\documentclass[]{aastex63}
\newcommand{\masy}{mas~yr$^{-1}$}
\newcommand{\kms}{km~s$^{-1}$}
\newcommand{\vtwod}{$v_{\rm 2D}$}
\newcommand{\vthreed}{$v_{\rm 3D}$}
\newcommand{\numGOSC}{455}
\newcommand{\numspecGOSC}{25}
\newcommand{\numBSN}{267} 
\newcommand{\msun}{M$_{\odot}$}

\newcommand{\numgroupthree}{215} 
\newcommand{\deltaPA}{$\Delta_{\rm PA}$} 

\shorttitle{Kinematics of Bowshock Stars}
\shortauthors{Kobulnicky \& Chick}

\begin{document}

\title{KINEMATICS OF THE CENTRAL STARS POWERING BOWSHOCK NEBULAE 
AND THE LARGE MULTIPLICITY FRACTION OF RUNAWAY OB STARS}

\correspondingauthor{Henry A. Kobulnicky}
\email{chipk@uwyo.edu}

\author[0000-0002-4475-4176]{Henry A. Kobulnicky}
\affiliation{Department of Physics \& Astronomy \\
University of Wyoming \\ 1000 E. University \\
Laramie, WY 82070, USA}

\author[0000-0002-8818-6780]{William T. Chick}
\affiliation{Department of Physics \& Astronomy \\
University of Wyoming \\ 1000 E. University \\
Laramie, WY 82070, USA}

\begin{abstract}

OB stars powering stellar bowshock nebulae (SBNe) have
been presumed to have large peculiar velocities.  We
measured peculiar velocities of SBN central stars to
assess their kinematics relative to the general O star
population using $Gaia$ EDR3 data for \numBSN\
SBN central stars and a sample of \numGOSC\ Galactic O stars to
derive projected velocities \vtwod.  For a subset
of each sample we obtained new optical spectroscopy to
measure radial velocities  and identify multiple-star
systems.  We find a {\it minimum} multiplicity fraction of
36$\pm$6\% among SBN central stars, consistent with
$>$28\% among runaway Galactic O stars.   
The large multiplicity fraction among runaways
implicates very efficient dynamical ejection rather  than
binary-supernova origins.   The median \vtwod\ of SBN
central stars is \vtwod=14.6 \kms, larger than  the median
\vtwod=11.4 \kms\ for non-bowshock O stars.  Central stars
of SBNe have a runaway (\vtwod$>$25 \kms) fraction of
24$^{+9}_{-7}$\%, consistent with the 22$^{+3}_{-3}$\% for
control-sample O stars. Most (76\%) of SBNe central stars
are not runaways.  Our analysis of alignment (\deltaPA)
between the nebular morphological and \vtwod\ kinematic
position angles reveals two populations: a highly aligned
($\sigma_{PA}$=25\degr) population that includes stars with
the largest \vtwod\ (31\% of the sample) and a random
(non-aligned)  population (69\%).  SBNe that lie within or
near \ion{H}{2} regions comprise a larger fraction of this
latter component than SBNe in isolated environments,
implicating localized ISM flows as a factor shaping their
orientations and morphologies. We outline a new conceptual
approach to computing the Solar LSR motion, yielding 
[U$_\odot$, V$_\odot$, W$_\odot$]= [5.5, 7.5, 4.5] \kms.

\end{abstract}

\section{Introduction} \label{sec:intro}
\subsection{Bowshock Nebulae and their Central Stars}

Stellar bow shock nebulae (SBNe) constitute a distinctive
class of arcuate interstellar nebulae flanking a central
star traveling supersonically through the interstellar
medium, first identified in association with $\zeta$~Oph and
LL Ori \citep{Gull1979}.  These nebulae contain swept-up
stellar wind material and trapped ISM at a circumstellar
distance  where the momentum flux of the ISM balances that
of the stellar wind \citep{Wilkin1996, Wilkin2000}. The
compressed and heated material of the nebula is most easily
seen in the infrared continuum, though some SBNe have
detectable emission lines including H$\alpha$ or \ion{O}{3}
\citep{Gull1979, Brown2005}. As more SBNe have been
discovered with advances in angular resolution and sky
coverage of infrared surveys, these nebulae have generated
increasing interest as laboratories to explore the
properties of both the central star and the ISM interactions
forming the nebular structure. SBNe and SBN-like objects
surround a wide variety of stars including red supergiants
\citep{Noriega-Crespo1997, Gvaramadze2014}, pulsars
\citep{Wang2013}, externally ionized proplyds
\citep{Bally1998}, and even an A dwarf  \citep[$\delta$ Vel,
][]{Gaspar2008}. \citet[][hereafter Paper I]{Chick2020}
conducted a spectroscopic survey of 84 cataloged 
SBN-candidates \citep[][hereafter K16]{Kobulnicky2016} and
determined that  $>$95\%  were O and early B stars, in
agreement with theoretical expectations  \citep{Weaver1977,
MacLow1991, Comeron1998}.

Hydrodynamical simulations indicate that the density of the
ambient ISM and the velocity differential between the ISM
and the central star are both significant factors in the
creation of an observable SBN \citep{Comeron1998, 
Meyer2014, Meyer2016}. Observable SBNe likely require
ambient interstellar densities of $\gtrsim$1--5 cm$^{-3}$
\citep{Kobulnicky2018, Henney2019}, making it  less likely
that SBNe can form or be detected inside  \ion{H}{2} regions
that efficiently heat and expel the ISM material
\citep{Hills1980, Goodwin2006, Dale2012} or in other
rarefied environments.  The star-ISM velocity differential
may be provided by the motion of the star
\citep{vanBuren1988, vanBuren1990, MacLow1991, vanBuren1992}
or by a bulk ISM flow impinging upon a stationary star 
\citep[i.e., ``in-situ'' bowshocks;][]{Povich2008}, and
perhaps a combination of both. Alternatively, the stellar
wind momentum flux may be supplemented (or even dominated)
by stellar radiation pressure \citep{vanBuren1988},
particularly around late-O stars in the ``weak wind regime''
\citep{Ochsendorf2014a, Ochsendorf2014b, Henney2019}. The
diverse range of interaction types has given range to an
equally diverse nomenclature of SBN subtypes, one still in
the process of being standardized. Some authors
\citep{Henney2019} have suggested three major subtypes:
traditional ``bow shocks'' where a thick shell of gas and
dust forms about a dust-free shocked wind-supported (or
radiation-supported)  inner layer, ``bow waves'' where a
thin shell of unshocked gas and dust forms due to the star's
radiation pressure acting on the dust and a gradual
deceleration by gas-dust drag, and ``dust waves'' where the
gas and dust decouple and a thin dusty shell forms supported
by radiation pressure but ISM gas continues to flow.
Variations and hybrids of each of these models have been
proposed as well \citep{vanMarle2011}. Throughout this work,
we will refer to all of these arc-shaped nebulae as SBNe,
acknowledging that this sample may contain some or all of
these physically distinct subtypes. \citet{Henney2019}
analyzed the 20 SBNe from \citet{Kobulnicky2018} and found
15 to be bow shocks, two likely bow waves, and three
ambiguous objects. 

The presumption that SBNe require supersonic (but not
hypersonic) star-ISM velocity differentials and the need to
form in denser regions of the ISM than their natal clusters
suggests that ``runaway'' stars \citep{Blaauw1961} are good
candidates to produce SBNe. Runaways exhibit large peculiar
velocities, having been ejected from their natal clusters
via N-body dynamical interactions \citep{Poveda1967,
Spitzer1980} or by supernovae in binary systems
\citep{Zwicky1957, Blaauw1961}. Indeed, K16 found 70\% of
their SBN central stars were in ``isolated'' environments
away from \ion{H}{2} regions. A minority of sources (7\%)
were co-located on the sky with \ion{H}{2}  regions. 
Although the threshold for runaway velocities varies by
author, most studies agree on velocities between 28 \kms\
\citep{Tetzlaff2011} and 40 \kms\ \citep{Blaauw1961}, with
most modern works adopting a three-dimensional peculiar
velocity threshold of 30 \kms\ \citep{Gies1986}. 
Additionally, most  runaways are OBA-type stars
\citep{Tetzlaff2011, delaFuenteMarcos2018}, though this may
be a selection effect due to their large luminosities. While
runaways provide a seemingly natural solution to both the
ISM density and velocity differential requirements to form
an SBN, it is clear that not all central stars supporting
SBNe are runaways and not all runaways support observable
SBNe. Some studies have found SBNe that are aligned with ISM
flows emanating from  \ion{H}{2} regions, suggesting the
presence of an in-situ SBN \citep{Gull1979, Povich2008}.
Similarly, \citet{Bodensteiner2018} found candidate SBNe
that show significant misalignment between their nebular
morphological  axis and their stellar proper motion vectors.
\citet{Peri2012, Peri2015} searched archival infrared images
of known OB runaway stars and found that only 15\% produced
SBNe observable in the infrared continuum. Based on these
works, it appears that detectable SBNe require a specialized
set of circumstances.  Being a runaway OB star alone is not
sufficient. 

\subsection{Runaway Stars and their Origins}

Observations indicate that runaway and ``walkaway'' stars
(peculiar velocities $<$30 \kms) comprise the
majority of {\it field} OB stars---those lying outside
known \ion{H}{2} regions and young clusters. 
\citet{deWit2005} found that only 4\% of field stars could
not be associated known associations and star forming
regions, implying that very few form in isolation. While
some additional OB stars may have formed in isolated
environments \citep[][]{Oey2018},  field and cluster OB
stars are statistically indistinguishable, except that
runaway O stars have later spectral types than those in
clusters \citep{vandenBergh2004}. It is likely that isolated
formation of O stars, if possible, is extremely uncommon. 

Kinematic investigations O stars have yielded significantly
different runaway fractions depending on the particular
observational selection criteria (e.g., definition of
runaway velocity threshold) or theoretical modeling
approach. Even studies that draw samples of stars from the
same catalogs can generate disparate results. For example,
based upon the radial velocities of stars from the catalogs
of \citet{Rubin1962} and \citet{Cruz-Gonzalez1974},
\citet{Stone1979} used a 40 \kms\ velocity threshold to find
an OB runaway fraction of 49$\pm$10\%, whereas
\citet{Gies1986} used a 30 \kms\ velocity threshold to find
a runaway fraction of 10--25\% for O stars and 2\% for early
B stars. The majority of studies identify runaways based
upon peculiar radial velocities or peculiar transverse
velocities alone.  Few studies incorporate all three
velocity components.  Assuming the distribution of stellar
velocities is isotropic, studies based upon the two
components of proper motion instead alone will underestimate
the true runaway fraction by 22\% \citep[i.e., 
$\sqrt{3/2}$;][]{Lamb2016}. By a similar reasoning, studies
based upon radial velocity alone may underestimate the true
fraction by as much as 73\% (i.e., $\sqrt{3/1}$).

In the seminal work on runaway O stars, \citet{Blaauw1961}
adopted a 3D runaway threshold of 40 \kms\ to find a runaway
fraction of 1.5\% for early-B stars, 2.5\% for B0 stars, and
21\% for late-O stars.  \citet{Stone1979} reanalyzed
archival Yerkes Observatory plates to measure the 2D
peculiar velocities of 47 O4--B2 stars. Using a
runaway threshold of 40 \kms\ they found an OB runaway
fraction of 49$\pm$10\%, though \citet{Gies1986} note that
this sample may have been biased toward runaways due to the
magnitude cut preferentially selecting stars with low
extinction. \citet{Gies1986} observed a population of 36 OB
stars selected based on previous identification as
high-radial-velocity sources complete to $V$=8 mag and as
faint as $V$=11 mag. After applying a statistical correction
to their observations to account for the sampling bias, they
determined a runaway fraction of 2\% for early-B and 10\%
for O stars. \citet{Stone1991} fit a bimodal distribution to
the observed velocities of O stars to find raw runaway
fractions of 17.4\%, 13.4\%, 9.5\% for 30, 35, and 40 \kms\ 
thresholds, respectively. After applying a statistical
correction for incompleteness the corrected O star runaway
fraction was 46\% \citep{Stone1991}.  \citet{deWit2005}
examined the O stars identified as field objects by
\citet{Mason1998} and found 22 of the 43 (51\%) to be
runaways based upon their 3D peculiar motions. It should be
noted that field populations would be expected to have
higher velocities due to their separation from stellar
associations and clusters. \citet{Lamb2016} identified a
sample of 374 ``field'' OB stars in the Small Magellanic
Cloud. Deriving peculiar radial velocities based upon the
radial velocities of nearby gas clouds, they found a lower
limit field OB star runaway fraction of 11.3$\pm$2.2\%,
though as noted above, they estimate the true runaway rate
may be as much as twice their reported rate if proper
motions were included. Based upon a statistical analysis of
Gaia DR1 \citep{Gaia2016b} proper motions of Galactic O-Star
Catalog stars (version 4.1), \citet{Maiz2018} found a
runaway fraction of 6.7$\pm$1.1\%. Seemingly, the only
consistent result between these studies is a disjunction
between the runaway fractions of O and B stars, with O stars
being much higher. These disparate results beg for
reexamination with the improved proper motions of $Gaia$
EDR3 data and radial velocities.

Despite the large dispersion in observed runaway fraction,
these observational studies conflict with simulations, which
predict very low fractions. Two mechanisms for generating
runaway stars have been proposed: the binary supernova
scenario (BSS), where stars with close binary companions are
ejected when the more massive component under goes an
asymmetric core-collapse supernova, and the dynamical
ejection scenario (DES), where stars are ejected through
N-body interactions. Hybrids models have also been proposed
to explain the isolated systems of \citet{deWit2005} through
the dynamical ejection of a binary system which then
undergoes the BSS.  Another hybrid idea involves a
supernova in a triple system that ejects a high-velocity
close binary \citep{Gao2019}. Studies of the DES mechanism
generally find higher proportions of stars achieving runaway
velocities and higher fractions of multiplicity among those
runaways than the BSS mechanism. 

Investigations of the efficacy of the BSS generally rely
upon population synthesis models, simulating the evolution
of (typically isolated) massive binaries. Such simulations
appear highly dependent on their adopted initial model
parameters, e.g., initial binary fractions, mass ratios,
orbital periods, and whether the simulation code includes
mechanisms such as companion mass exchange or N-body
dynamical interactions. Assuming an initial binary fraction
of 80\%, \citet{DeDonder1997} found that 5--8\% of O stars
have runaway velocities $>$30 \kms, and fewer than one-third
of those runaway systems are binaries. Modeling only binary
systems, \citet{Eldridge2011} found a runaway fraction of
approximately 5\% for O stars and slightly lower for B
stars. \citet{Renzo2019} and \citet{Evans2020} modeled the
evolution of isolated binary systems and found BSS-induced O
star runaway fractions of 0.5$^{+1.0}_{-0.4}$\% and
0.5--2.0\%, respectively. It should be noted that
\citet{DeDonder1997}, \citet{Renzo2019}, and
\citet{Evans2020} do not consider the additional effect of
dynamical ejections in their studies, focusing instead on
isolated systems which likely contributes to their lower
runaway fractions. \citet{Eldridge2011} acknowledged the
likelihood that the DES contributes to the runaway
population but predicted only modest increases in the
runaway fraction over their strictly BSS predictions. 

N-body simulations of stellar clusters predict that
dynamically ejected runaways comprise a few percent of the
general runaway O star population, similar to the
contribution of the BSS. \citet{Poveda1967} modeled 5- and
6-body star clusters analytically and found that 2--17\% of
stars were ejected with velocities in excess of 35 \kms.
Using N-body simulations of 10$^{3.5}$ \msun\ and 10$^{4.0}$ \msun\
stellar clusters, \citet{Oh2015} found about 30\% of the
ejected O stars achieve runaway velocities, producing a
runaway fraction of 5\% along all O stars. One of the
few studies that approaches the observed rate of runaways is
the DES simulations of \citet{Perets2012} that predicted O
runaway fractions of 5--20\% and B-star runaway fractions of
one-third to one-sixth that of the O star fraction. 

It appears likely that both the BSS and DES contribute to
the population of runaways. Indeed, some observational
studies have identified runaway star candidates from
clusters that are too young to have experienced their first
supernova \citep[e.g.,][]{Gvaramadze2008, Roman-Lopes2013,
Drew2018} making DES the only likely mechanism for producing
these systems. Similarly, studies have identified
high-proper-motion star+pulsar pairs with anti-parallel
velocity vectors that suggest a common co-located origin,
making them probable BSS-generated runaways
\citep{Hoogerwerf2000, Hoogerwerf2001}. Given that each
mechanism generally produces runaway fractions of a few to
several percent in simulations, combinations of the two
mechanisms can only reproduce the low end of the observed
runaway fractions, i.e., $<$10\%.

\subsection{Goals of this Investigation}

Here and Paper I we present the results of
complementary studies to \citet{Peri2012},
\citet{Peri2015},  and \citet{Bodensteiner2018}. Whereas
those works began with catalogs of runaways and bright OBA
stars, respectively, and searched for SBNe, we begin with
the comprehensive K16 and supplementary
\citet{Jayasinghe2019} catalogs  of SBNe to determine their
spectral types (Paper I) and kinematic properties (this
work). With the high-quality measurements from the  Gaia
Early Data Release 3 (EDR3) \citep{Gaia2016a, Gaia2018}, it
is possible to locate individual stars within the Galactic
Plane to distances of nearly 8 kpc, greatly extending the
accessible volume over previous stellar catalogs such as
Hipparcos \citep{vanLeeuwen2007} or Tycho-2 \citet{Hog2000}.
Previous measurements of the velocities of SBNe
\citep{vanBuren1995, Bodensteiner2018} have measured proper
motions without accounting for the Galactic rotation curve 
and often without correcting for solar peculiar motion
(e.g., K16). While this is reasonable approximation when
within the solar neighborhood, at distances of $>$few kpc
these corrections can be on the order of 10--30 \kms,
similar to the  runaway velocity threshold. In this work, we
account for both the solar peculiar motion and the expected
contribution of the Galactic rotation curve to calculate
peculiar velocities  within each star's local standard of
rest.

The majority of stars in the K16 catalog lack radial
velocity measurements, precluding the calculation of true 3D
peculiar motions. However, simulations of SBNe suggest that
their arcuate shapes become increasingly reniform (circular)
as the inclination angle increases beyond $\approx$45\degr\
from transverse \citep{Acreman2016, Meyer2016}. As SBNe in
the K16 catalog were identified on the basis of an arcuate
morphology, this suggests that, if the star-ISM differential
is created by the stellar peculiar velocity, the largest
components of the peculiar velocity will be transverse and
not radial. For this reason we adopt and defend the
proposition that the  two-dimensional peculiar velocity
(\vtwod) calculated from proper motions is a reasonable
approximation of the full three-dimensional velocity
(\vthreed) for most SBN central stars.

In this work, we calculate \vtwod\ for a sample of SBN
central stars and a comparison sample of O stars from
Galactic O-Star Catalog \citep[][ GOSC]{Maiz2013, Maiz2016}
to determine their velocity distribution and runaway
fractions.  As a comparison sample, the GOSC contain massive
stars earlier than B0 selected for completeness without
regard to other properties, whereas the SBN sample is
selected on the basis of infrared nebular morphology. 
Furthermore, the GOSC  contains exclusively O stars while
the SBN sample contains both O and early B
stars---predominantly the latter.  If the kinematic
properties of O and early B stars differ systematically
(e.g., as suggested by \citet{Kobulnicky2016} and
\citet{Banyard2022} with regard to O stars having a higher
multiplicity fraction than B stars), then the comparisons
between the SBN and GOSC samples may be regarded as
approximate but imperfect.

Our goals include understanding the origins of runaway
stars,  the physical mechanisms responsible for production
of SBNe, and the role of multiplicity among SBN stars and O
stars.   If the SBN central stars show an excess of runaways
relative to a general O star sample, this would support the
idea that stellar peculiar velocities are responsible for
the star-ISM velocity differential. However, if the peculiar
velocities are the same as the general O-star population,
this may indicate that either ``in situ'' shocks comprise a
larger portion of the SBNe population or that ISM density is
a more important factor in the formation of an SBNe than the
velocity differential.  If runaway stars  are predominantly
single relative to the field O population then the  BSS
mechanism is the favored one.  However, if runaways
are commonly multiple-star systems, then the DES mechanism
must be responsible for their large velocities. 

Section 2 describes the selection of sample sources, the
methods and assumptions of the kinematic computations, and
spectroscopic reduction techniques for GOSC runaway O stars
and SBN central stars observed.   Section 3 presents results
on the GOSC O-star runaway comparison  sample, including the
distribution of \vtwod, runaway fraction, and multiplicity
fraction for a subsample of \numGOSC\ GOSC sources. Section
4 presents two-dimensional and three-dimensional velocities
of the SBN central stars sample, runaway fractions, and the
distribution of alignments between the stars' velocities and
the orientation of their nebulae. In Section 5 we conclude
with a discussion of these results and implications for
production of SBNe and runaway stars.

\section{Kinematic Analysis and New Spectroscopic Data}
\subsection{Source Selection}

The stellar sources for this work were selected from the K16
catalog of 709 morphologically identified SBNe candidates
and their central stars, supplemented by an
additional\footnote{The objects were selected by senior
author H.A.K. using the same morphological criteria as in
K16 from among the 311 \citet{Jayasinghe2019} additional
candidate stars not  contained in the K16 catalog.} 187  
citizen-science-identified stars from
\citet{Jayasinghe2019}.  We extracted $Gaia$ EDR3
astrometric quantities for these 896 sources as matched by
position using a 1.5\arcsec\ search radius---twice the
angular resolution of the mid-IR images that comprise the
search dataset---from the position of the mid-infrared point
source identified as the probable central star of each
nebula. Positional matching resulted in 797 bowshock central
stars having a viable EDR3 counterpart. Those without a
match are likely objects where the central star is too faint
at optical wavelengths to appear in the EDR3.  In 19
instances positional matching resulted in two candidate
central stars. For the majority of such cases the ambiguity
was resolved by selecting the source with the optical
magnitude most consistent with the infrared-identified point
source IR magnitudes.  Of the  stars having an EDR3 entry,
only 582 have a physically meaningful positive parallax.  We
removed 78 stars having a renormalized unit weight error
parameter \citep[][RUWE]{Lindegren2018} $>$1.4, as these
have astrometric solutions showing larger-than-expected
residuals---possibly as a consequence of
binarity---resulting in less reliable proper motions.
Finally, we retained only the stars having parallax and
proper motion measurement:error ratios greater than 4:1,
leaving \numBSN\ objects having well-measured astrometric
parameters.  The 16th/50th/84th percentiles of the $Gaia$
G-band magnitudes of selected stars are 9.5/12.6/14.8, so
the majority of the sources are considered bright targets.

We created a field O star (non-bowshock)
comparison sample from the 590-source GOSC, as compiled in
\citet{Maiz2016} using the same selection criteria as for
the SBN central stars, resulting in \numGOSC\ O stars
comprising the GOSC comparison sample.

\begin{figure}[ht!]
\fig{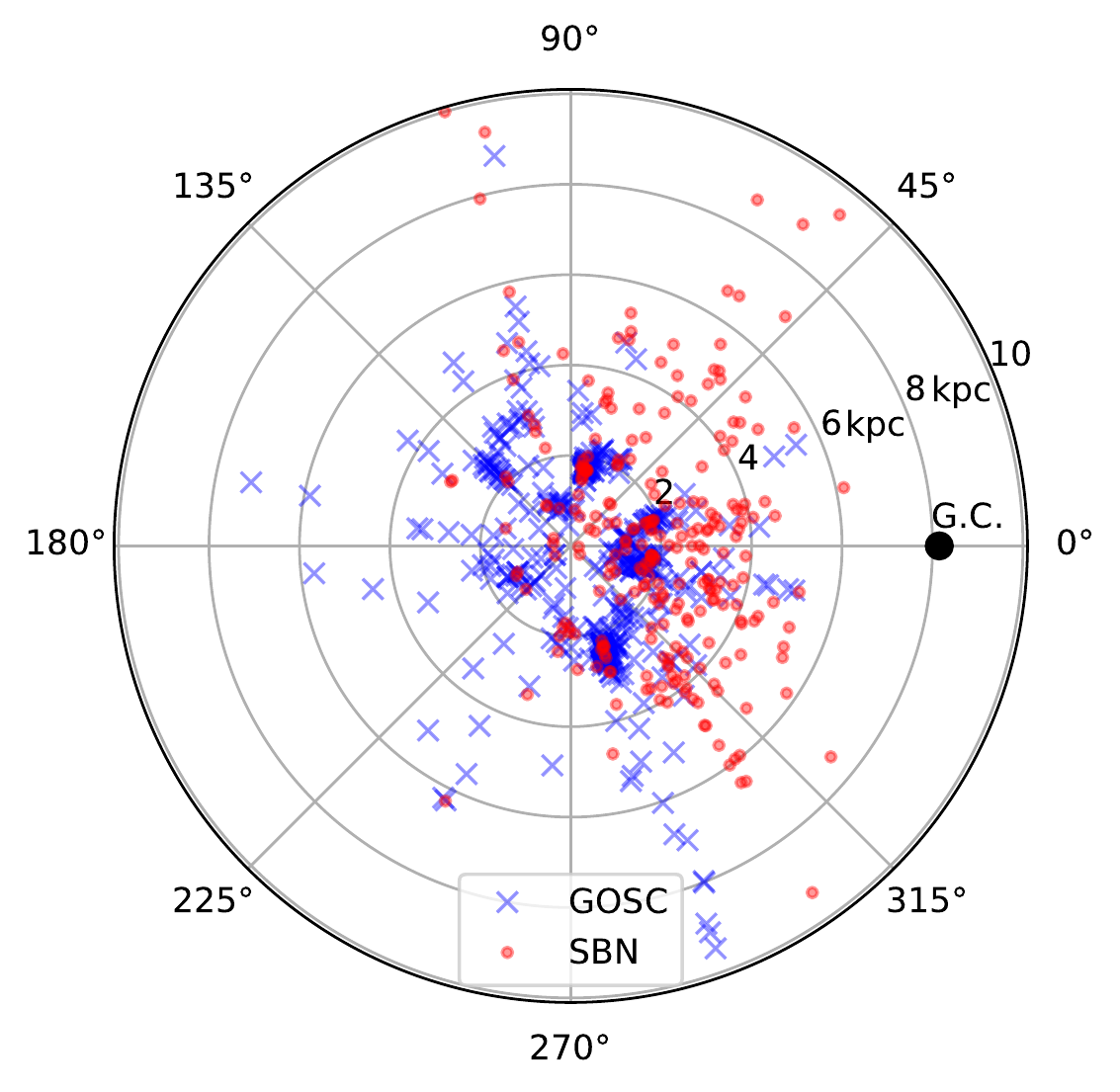}{5in}{}
\caption{Polar representation of the Galactic locations of
GOSC O stars ({\it blue x's}) and SBN central stars ({\it red dots}), 
with the Sun at center and Galactic Center toward the right.  
 \label{fig:gal}}
\end{figure}

Figure~\ref{fig:gal} plots the Galactic locations of the
\numGOSC\ GOSC O stars ({\it blue x's}) and \numBSN\ SBN
central stars ({\it red dots}) in each sample using a polar
representation with the Sun at center and Galactic Center
toward right.  The vast majority of the SBN central stars
lie in the inner Galaxy, with a comparable number in the
first quadrant (112 objects) and fourth quadrant (117
objects).  Nearly all SBN objects lie within 1\degr\ of the
Galactic Plane. The mean distance is larger for SBN central
stars (3.3 kpc) than for GOSC O stars (2.3 kpc), a likely
consequence of the  infrared selection criteria for the
former and optical selection  criteria for the latter.  

\subsection{Kinematic Calculations}

We computed the Galactic position and peculiar  velocity 
(\vtwod\ from the two proper motion components)  for each
source relative to its local standard of rest based upon the
EDR3 proper motions and parallax-based distances.
Uncertainties on parallaxes and proper motions are, on
average, factors of 3--5 smaller in the EDR3 compared to the
earlier DR2 dataset. We adopted  inverse parallax values for
the distances. The calculation of  velocities is based upon
the matrix equations of \citet{Johnson1987} in conjunction
with an adopted solar motion relative to the Local Standard
of Rest (LSR) and an adopted Galactic rotation curve. For a
complete description of this technique and the associated
matrix transformations, see Appendix A of
\citet{Randall2015}. To transform from  equatorial to
Galactic coordinates, we adopted the \citet{Reid2004}
Galactic Center coordinates ($\alpha_{\rm GC}$ =
17$^h$:45$^m$:37\fs224$s$, $\delta_{\rm GC}$ =
$-$28\degr:56\arcmin:10\farcs{23}) and Galactic North Pole
coordinates ($\alpha_{\rm GNP}$ = 12$^h$:51$^m$:26\fs282,
$\delta_{GNP}$ = 27\degr:07\arcmin:42{\farcs}01). We adopted
a Solar Galactocentric distance of 8.15 kpc \citep{Reid2019}
and the \citet{Irrgang2013} Model II Galactic rotation
curve.  The choice of rotation curve makes very little
difference to the calculated  velocities because all modern
rotation curves are nearly flat within a few kpc of the Sun
where most of our sample stars lie. The calculation of
peculiar stellar velocities, however, is somewhat sensitive
to the adopted Solar motion relative to the LSR.  We tried
both the solar peculiar motions of \citet{Schonrich2010},
[U$_\odot$, V$_\odot$, W$_\odot$]= [11.1$\pm$1.2,
12.2$\pm$2.1, 7.2$\pm$0.6] \kms\ and those of
\citet{Ding2019} which are based on Gaia DR2 stellar
kinematics,  [U$_\odot$, V$_\odot$, W$_\odot$]=
[8.63$\pm$0.64, 4.76$\pm$0.49, 7.26$\pm$0.36] \kms.  We
found neither to be satisfactory.  Motivated by the reasons
and analysis described in the  Appendix we adopted
[U$_\odot$, V$_\odot$, W$_\odot$]= [5.5$\pm$1, 7.5$\pm$1,
4.5$\pm$1] \kms.   These values are within the uncertainties
and plausible ranges of many historical determinations
\citep[e.g., see summary table of][]{Ding2019}.  

We estimated the uncertainties on the peculiar velocities
using a 1000-iteration Monte Carlo analysis, taking into
account the EDR3 uncertainties on parallax and the two
proper motion components, along with the uncertainties on
the solar motion relative to the local standard of rest.
This method produces distributions of peculiar velocities in
U, V, and W. We then added in quadrature the  uncertainties
of each peculiar velocity distribution to determine the
uncertainty on \vtwod. We did not consider systematic
uncertainties on the EDR3 motions, as these are estimated to
be $\lesssim$0.004 \masy, negligible compared to the random
uncertainties and small compared to the earlier Gaia DR2
systematic uncertainty of $\sim$0.010 \masy
\citep{Gaia2021}.  Likewise, the systematic error on the
EDR3 parallaxes is estimated at $\sim$0.010 mas, negligible
for the purposes of this work \citep{Lindegren2021}.  
Uncertainties on EDR3 parallaxes may be up to a factor of
two larger than the published values as a complex function
of location on the sky, color, and magnitude
\citep{Lindegren2021, Maiz2021, Cantat2021}, but we do not
consider these owing to the evolving understanding of $Gaia$
systematics as of this writing.  Proper motions of bright
(G$<$13 mag) stars may also be systematically rotated with
respected to fainter targets \citep{Cantat2021}, so we
performed an analysis with and without their recommended
corrections using their code and concluded that these are 
insignificant to the conclusions of this work.   

For those sources with two or more radial velocity
measurements from our spectroscopic survey, we calculated a
3-dimensional peculiar velocity (\vthreed) using the
weighted mean radial velocity measurement and adopting  the
RMS of those measurements as the associated uncertainty.
This mean is used to minimize the effects of potential
radial velocity deviations from binary (or higher-order
multiple) systems, both identified and unidentified.

Adding signed quantities in quadrature (i.e., the two
components of proper motion)   always contributes positively
to the final value. We correct for this positive bias by
subtracting in quadrature the velocity uncertainties
($\sigma_{\rm i}$) from the calculated velocities
($v\prime_{\rm i}$) for each star,

\begin{equation} 
v_{\rm i}= \sqrt{v\prime_{\rm
i}-\sigma_{\rm i} }. 
\end{equation}

\noindent For the selected sources with small fractional
uncertainties, this bias correction is small relative to the
uncorrected velocities. The median positive bias correction
is 1.7 \kms\ for the GOSC stars and 1.8 \kms\ for the SBN
stars.  In order to test and validate our code, we
replicated the \citet{Sperauskas2016} analysis of the peculiar
motions of 1088 K--M stars in the solar neighborhood using
their set of assumptions and Galactic rotation curve. The
average absolute difference between their calculated
peculiar  velocities and ours is 1.1$\pm$1.0 \kms.

\subsection{Spectroscopic Observations of GOSC O Stars}

In addition to the optical spectra acquired previously on 84
SBN stars for Paper I, we obtained spectra of \numspecGOSC\
stars from the GOSC, selected on the basis of having large
\vtwod\ (in excess of 25 \kms), high source brightness
(broad range of 7.2$\lesssim$$V$$\lesssim$11.2 mag), and positive
declinations suitable for observation from the northern
hemisphere.  No sharp magnitude cut was applied so that the
sample is not biased toward  more luminous (hence, binary)
stars. These sources also met the astrometric selection
criteria described above. We obtained 119 new optical
spectra using the Wyoming Infrared Observatory (WIRO) 2.3
meter and Apache Point Observatory (APO) 3.5 meter
telescopes. Each star in this sample was observed on at
least three separate nights, with thirteen stars observed
more than five nights. The interval between observations
varied between one and 84 days with a mean of 30 days and a
median of 32 days. The mean time baseline from first to last
observation was 107 days with a median of 118 days. All but
one star had a minimum baseline of 57 days with 17 of the
\numspecGOSC\ stars having a baseline longer than 90 days. 

We acquired 81 spectra on the nights of 2019 July 29, August
16, 28, and 19, and September 4 using the WIRO LongSlit
spectrograph. Observational methodologies for WIRO spectra
closely follow those described in Paper I. During each
observation, sources were positioned within a
1\farcs{2}$\times$120\arcsec\ slit oriented north-south. A
2000 line mm$^{-1}$ grating yielded a spectral resolution of
1.26 \AA\ FWHM ($R$$\approx$4000) with a spectral coverage of 5400--6800 \AA.
Exposure times ranged from 1$\times$30 s to 2$\times$300 s,
depending on source brightness and seeing. CuAr arc lamp
exposures were taken before and after each exposure,
yielding wavelength solutions with a typical RMS of 0.015
\AA, or 0.8 \kms\ at 5800 \AA.  The wavelength calibration
is estimated to be precise to 6 \kms.

We acquired 38 spectra using the Double Imaging Spectrograph
(DIS) on the APO 3.5 meter telescope over seven
nights: 2019 May 24, 2019 July 3 and 9, 2019 September 4 and
25, and 2019 October 22 and 29. The 1200 line mm$^{-1}$
grating of the red spectrograph arm yielded a reciprocal dispersion
of 0.58 \AA\ pix$^{-1}$ over the 5700--6900 \AA\ wavelength
range. Depending on source brightness, exposure times ranged
from 1$\times$20 s to 2$\times$120 s. The instrument HeNeAr
arc lamp supplied wavelength calibration to an RMS of 0.06
\AA. Instrument rotation during the night produces
wavelength shifts of up to 0.3 pix, which were removed using
periodic HeNeAr arc lamp exposure such that the wavelength
solutions are estimated to be precise to about 5 \kms.

Spectra were extracted and reduced using standard longslit
techniques within IRAF \citep{Tody1986}. Flat-fielding was
performed using quartz lamp dome flats for WIRO spectra and
internal quartz lamp flats for APO spectra. Where multiple
spectra were taken in a single target, nightly spectra were
combined into a single master spectrum yielding a final
signal-to-noise ratio (SNR) of at least 65 for APO spectra
and 80 for WIRO spectra in the vicinity of the stellar
\ion{He}{1} 5786 \AA\ line used to measure radial
velocities. Final spectra were continuum normalized and
Doppler corrected to the Heliocentric velocity frame. Radial
velocities were extracted using the methodology described in
Paper I. Briefly, the Heliocentric velocity   for each
spectrum was re-zeropointed by shifting it  (typically 3--15
\kms) to place the centroids of the interstellar Na I D
$\lambda\lambda$5890, 5896 lines at the mean velocity for
the ensemble of each star. Using a fitting code based around
the IDL routine MPFIT \citet{Markwardt2009}, we measured
radial velocities from each spectrum by fitting Gaussian
functions to the \ion{He}{1} 5786 \AA\ line. This fitting
code fixed the Gaussian width and depth to the mean from the
ensemble  after manually rejecting outliers and solved for
the best line center and uncertainty. This technique allowed
us to measure robust {\it relative} radial velocities but,
since our data lack radial velocity standards, these should
not be treated as absolute radial velocities due to
potential systematic offsets relative to the Heliocentric
frame. We estimate any such absolute systematic effects to
be small, $<$8 \kms, based on calibrations of other surveys
using the same instrument \citep{Kobulnicky2014}.

\section{GOSC Comparison Sample Results}
\subsection{Runaway Fraction of the General O Star Population}

\begin{figure}[ht!]
\fig{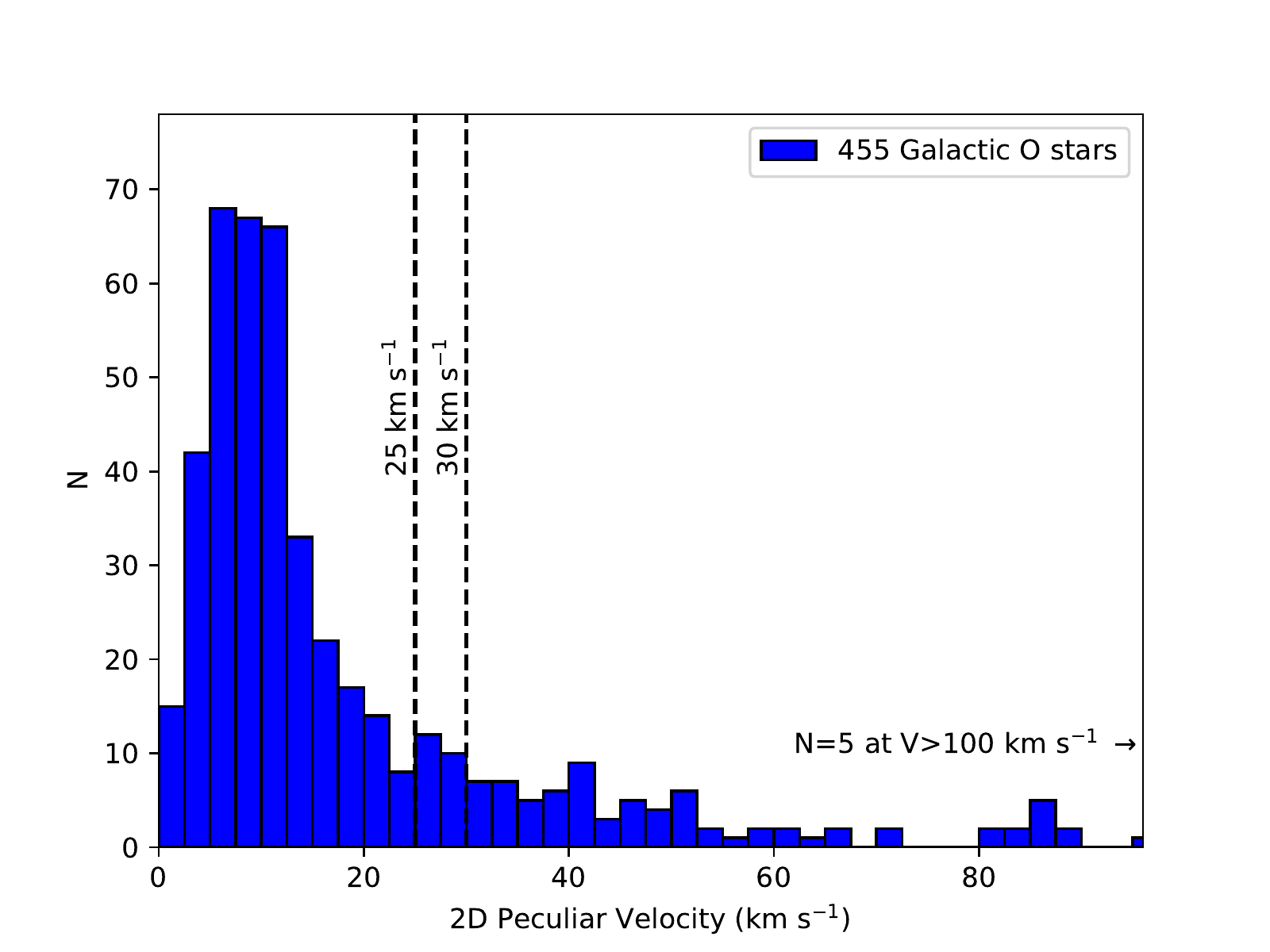}{5in}{}
\caption{A histogram bias-corrected 2D peculiar velocities of \numGOSC\ 
O stars from the Galactic O Star Catalog \citep{Maiz2016}. 
Vertical dashed gray lines indicate the \vtwod\ runaway thresholds 
of 25 and 30 \kms. \label{fig:GOSC2D}}
\end{figure}

Figure~\ref{fig:GOSC2D} presents a histogram of the
distribution of \vtwod\ for the \numGOSC\ GOSC stars, using
a bin width of 2.5 \kms. Vertical dashed lines mark the two
fiducial thresholds for runaways at \vtwod$>$25 \kms\ and
\vtwod$>$30 \kms, respectively. Five systems lie off the
right edge of the  plot at \vtwod$>$100 \kms. The
distribution in Figure~\ref{fig:GOSC2D} is similar to the
peculiar tangential velocities of \citet[][Figure
4]{Tetzlaff2011} having a peak near 8 \kms\  with a long
tail extending out to 100 \kms. \citet{Tetzlaff2011} do not
apply a positive bias correction so their distributions may
be biased to slightly larger velocities. The median/mean
velocity uncertainties are 1.8/2.6 \kms. 

Of the  \numGOSC\  GOSC sources, 81$^{+14}_{-8}$, have
\vtwod$>$30 \kms, yielding a runaway fraction of
16$^{+3}_{-1}$\%. For this work, we adopt a \vtwod\ runaway
threshold of 25 \kms\  as our primary runaway star
criterion, for two reasons. \citet{Tetzlaff2011} analyzed
7663 stars within 3~kpc  of the Sun and found the peculiar
tangential velocities (our \vtwod) could be described by a
two-component Gaussian distribution, with low- and
high-velocity components peaking at 7 and 22 \kms,
respectively. These low- and high-velocity distributions
intersect at 20 \kms,  with the low-velocity component
approaching zero amplitude at $\geq$25 \kms. Therefore,
adopting 25 \kms\ as our \vtwod\ threshold would select only
members of this high-velocity group. Furthermore, a
\vthreed\ threshold of 30 \kms\ corresponds to a \vtwod\
threshold 25 \kms\ ( $\sqrt{3/2}$$\approx$30/25). After
applying the positive bias correction to the data, our
\vtwod\ threshold of 25 \kms\ yields 102$^{+14}_{-15}$ of
\numGOSC\ GOSC O stars as runaways (22$^{+3}_{-3}$\%).

Our GOSC runaway fraction is notably higher than that found
by \citet{Maiz2018} who analyzed 594 GOSC stars (including
some previously unpublished stars) using Gaia DR1 proper
motions. However, \citet{Maiz2018} used a statistical
analysis based on 2D proper motions corrected for solar
motion, rather than a true peculiar velocity. This method
was used for computational simplicity and to avoid the use
of DR1 parallaxes which did not include bright sources and
had large relative uncertainties. The statistical cut for
their runaway candidates was empirically determined using
the results of \citet{Tetzlaff2011}. \citet{Maiz2018} found
49 runaway candidate O stars out of 594 sources, yielding a
runaway fraction of 8.2\%, considerably lower than our 
22$\pm$3\%, as well as lower than historical literature
values for O stars. Opting to use the \vtwod$>$30 \kms\
fraction of 17$^{+4}_{-1}$\% reduces but does not resolve
the inconsistency. One potential reason for their lower
runaway rate may be their analytical method being
insensitive to stellar distance. In our analysis 77 (17\%)
of the GOSC sources have distances greater than 3 kpc.  At
greater distances large transverse velocities have
proportionally smaller proper motions.  Because of their
focus on proper motion outliers, their analysis  may exclude
high-peculiar-velocity  stars at large distances. Of  the
\citet{Maiz2018} runaway candidates our analysis recovered
all 30 that met our selection criteria.  Our new field O
star runaway fraction of 22$\pm$3\% is consistent with most
previous observations of O stars as outlined in Section 1,
including the raw fraction found by \citet{Stone1991} and
the \citet{Blaauw1961}  O stars,  but substantially lower
than the \citet{Stone1991} corrected fractions of 46\% and
lower than the 55\% reported by \citet{deWit2005}.

\subsection{Multiplicity Fraction of Runaway GOSC Stars}

Table~\ref{tab:GOSCRunaway} lists the computed stellar
kinematic properties for each of the \numspecGOSC\
spectroscopically studied GOSC runaway candidates. Columns 1
and 2 list the GOSC catalog identification number and a
common alias, respectively. Column 3 is the EDR3 RUWE. 
Columns 4 and 5 are the calculated \vtwod\ and 1$\sigma$
uncertainty.  Column 6 is the mean Heliocentric radial
velocity of the 3--6 individual measurements, column 7 is
the average velocity uncertainty, and column 8 is the RMS of
the 3--6 radial velocity  measurements.  Column 9 gives 
probability $P$($\chi^2$, $\nu$)  for each star's set of
radial velocities being consistent with the null hypothesis
(no velocity variability) using a chi-squared test with
$\nu$  degrees of freedom, where $\nu$ is the number of
radial velocity measurements  minus one.  Sources with 
$P$($\chi^2$, $\nu$)$<$0.05 are inconsistent with the null
hypothesis of a single-star system and are interpreted as
candidate multiple-star systems.   Column 10 lists the
spectral type and luminosity class from the literature, and 
column 10 gives the dates of observation for each star in a
modified Heliocentric Julian Date format (HJD$-$2,458,000).

\begin{figure}[ht!]
\fig{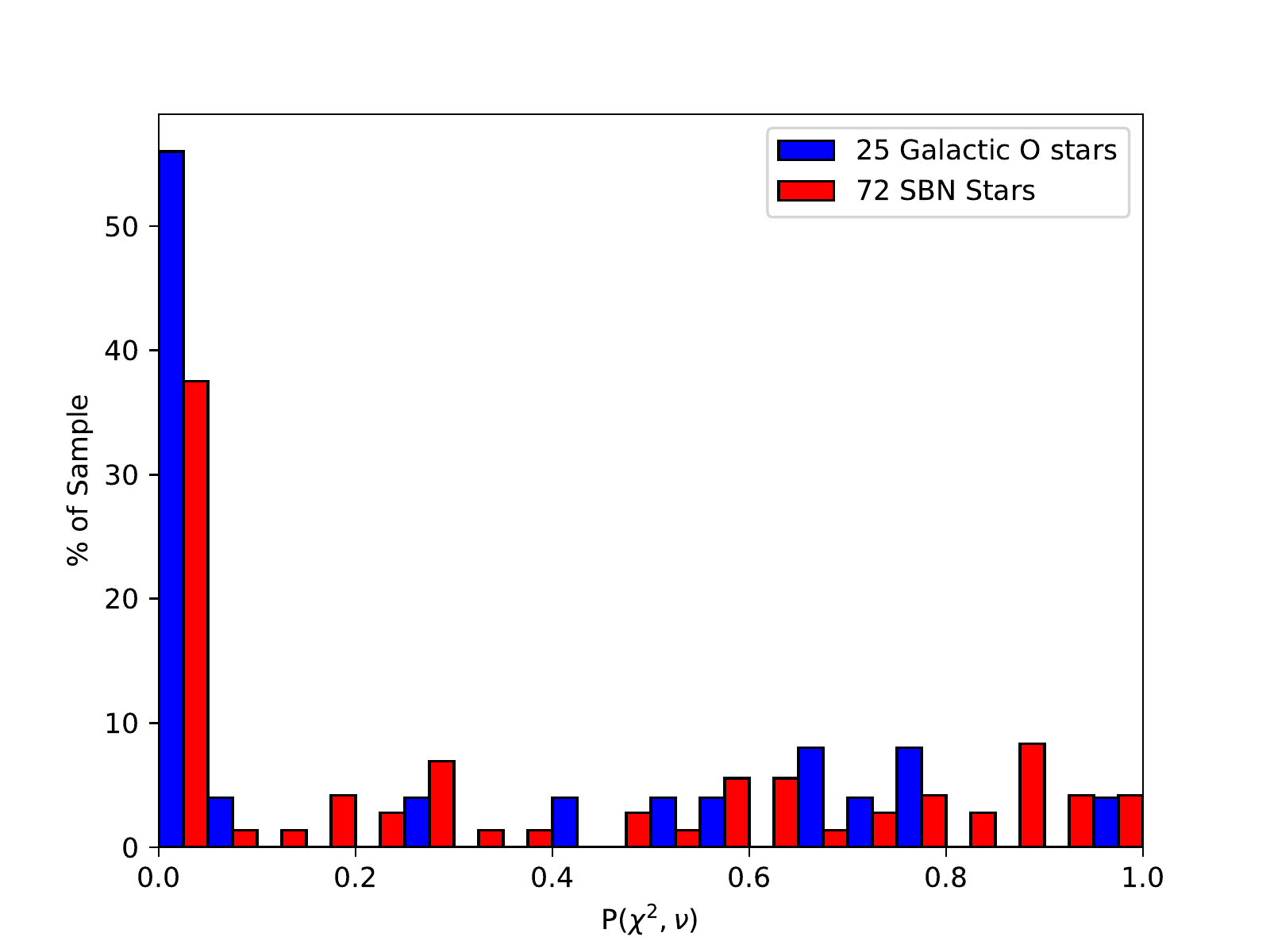}{5in}{}
\caption{A histogram chi-squared probability distribution for velocities of 25 
O stars from the Galactic O Star Catalog and 72 central stars of bowshock nebulae. 
\label{fig:histprob}}
\end{figure}

Figure~\ref{fig:histprob} displays a histogram comparing the
distributions of $P(\chi^2$, $\nu$) for the \numspecGOSC\ 
spectroscopically studied GOSC O stars  and the 72 SBN
central stars having multi-epoch radial velocity data
presented in Paper~I.  The distributions of P($\chi^2$,
$\nu$)  are similar for the two populations, showing a
roughly flat distribution as expected for a population of
single-star systems but with a sharp peak at $P$($\chi^2$,
$\nu$)$<$0.05 indicating a population of candidate
multiple-star systems. Of the 25 GOSC systems, 14
(56$\pm$16\%) have $P$($\chi^2$, $\nu$)$<$0.05 compared to 27
of the 72 SBN central stars (38$\pm$6\%). If we were to
consider only the 52 SBN stars meeting our astrometric
selection criteria, then  16 (31$\pm$7\%) have P($\chi^2$,
$\nu$)$<$0.05.   Assuming the average occupancy of those
bins with P($\chi^2$, $\nu$)$>$0.05 to be an estimate of the
false positive rate for each sample yields corrected
multiplicity numbers (fractions) of 13.4 GOSC stars
(53$\pm$14\%) and 25.6 SBN central stars (36$\pm$6\%). If we
consider only the O stars among the SBN central stars, 3 of
8 (38$\pm$22\%) show evidence of multiplicity. These results
show that both the full OB-star SBN sample and O-star SBN
 sample are consistent with the runaway GOSC O-star
 multiplicity fraction within 1$\sigma$.  The 
apparent multiplicity fraction of the GOSC and SBN samples are
formally consistent, despite the fact that early B stars
dominate  the SBN sample \citep{Chick2020}, and B stars have
been shown to exhibit a lower multiplicity fraction than O
stars \citep{Kobulnicky2014,Banyard2022}.  This bias should
augment the disparity in spectral type between the samples. 
The consonance between SBN subsamples suggests 
that SBN central stars, despite consisting mostly of early B
stars, may have larger multiplicity fractions than GOSC O
stars.  Or, this result could be interpreted to  mean that
binary systems are more effective in generating observable
stellar bow shock nebulae, since the SBN sample  is
constructed exclusively by the presence of these nebular
infrared features.

We considered the possibility that quasi-periodic line
profile variations attributed to non-radial pulsational
modes  and/or stochastic atmospheric fluctuations among
early type stars (rather than multiplicity) produce  
some of the observed variability and artificially
inflate the  multiplicity fraction of the GOSC sample. 
\citet{Fullerton1996} found that 23 of 30 O stars---and all
of the supergiants in their sample---exhibited significant
line profile variations in resolution
$R$$\approx$20,000--30,000 optical spectra.  Some of these
would be large enough to mimic the signature of binarity
seen in our $R$=4000 spectra.  However, in a spectroscopic
survey of 128 O and early B stars that yielded orbital
solutions for 48 systems using at least 12 observations per
system, only six stars were found to exhibit aperiodic line
variations attributed to atmospheric effects
\citep{Kobulnicky2014}.   All six were supergiants.  Among
the present sample, 7 of the 14  GOSC stars showing velocity
variability are supergiants.  If we conservatively suspect
all of the supergiants of being dominated by pulsational
 variations and remove them, this leaves 7 of 25
(28\%) as multiple-star candidates.  A more modest correction 
based on the \citet{Kobulnicky2014} survey suggests removing 
 2--3 velocity-variable supergiants as potential contaminants.   
These considerations lead us to adopt a {\it revised multiplicity
fraction of $>$28--48\% for GOSC runaway O stars.}  
  
All of these fractions should be considered {\it minimum}
multiplicity fractions  considering the short observational
time baseline for the GOSC spectroscopic observations.  The
data on the SBN central stars presented in Paper I, with its
longer observational time baselines, was able to identify
all five previously recognized spectroscopic binary sources.
Among the GOSC spectroscopic sample, five stars (GOSC IDs
27, 206, 234, 340, 365) are identified in the literature as
spectroscopic binaries, primarily by \citet{Pourbaix2004}.
Our spectroscopic program identified three of these (27,
206, and 365) as having $P$($\chi^2$, $\nu$)$<$0.05 but did
not identify the two other literature binaries. GOSC 340 (HD
192281) was originally identified as a binary system by
\citet{Barannikov1993} with a period of 5.48 days and a
semi-amplitude of 16.8$\pm$2.4 \kms\  (though
\citet{Maiz2019} note this has not been subsequently
confirmed). The average velocity uncertainty for this source
in our dataset is 13 \kms, of a similar magnitude as the
semi-amplitude. The orbital period of GOSC 234 (HD 15137) is
28.6$\pm$0.4 days \citet{McSwain2007}. The irregular
observational cadence of our study  was 56.0, 1.0, and 60.7
days between observations, approximately two periods. This
unfortunate alignment of cadence and orbital phase explains
why our data do not reveal its multiplicity.  Including
these two literature binaries would raise the raw
(uncorrected for line profile variability contamination) 
multiplicity fraction of our GOSC sample to 16 out of 25
sources---64\%---comparable to multiplicity fractions of
other OB star samples \citep{Kobulnicky2014,Banyard2022}. 
Our data reinforce the conclusion that multiple-star systems
are common, not just among OB stars in general, but among 
high-velocity runaways as well.  

\section{SBN Central Star Sample Analysis}

\subsection{Kinematics of the Central Stars of Stellar Bowshock Nebulae}

We divided the SBN central stars that met our astrometric
selection criteria into three subsamples. The first group is
the ``Multiple-Star Candidates'':  16 sources, selected on
the basis of having $P$($\chi^2$, $\nu$)$<$0.05.   The
second group is the ``Single Star Candidates'': 36 sources
having $P$($\chi^2$, $\nu$)$\geq$0.05.  There
are likely to be multiple-star systems in this group that
were not identified in our radial velocity survey. However,
the fraction of binaries in this sample will be lower than a
typical SBN central star sample. The third group, termed
``Group 3'', consists of the \numgroupthree\ sources that met our
astrometric selection criteria but do not have measured
radial velocities.   This sample is likely the most
representative of SBN sources in general based upon the
multiplicity cuts for the first two groups. 

Table~\ref{tab:BSNkinematics} provides identifications and
tabulated astrometric data for the \numBSN\ SBN central
stars meeting our astrometric selection criteria. Column 1
is either the K16  catalog identification number from
\citet[][numbers 1--709]{Kobulnicky2016} or an
identification index that includes the 187 additional stars
drawn from \citet[][numbers 710--896]{Jayasinghe2019}.
Column 2 is a common  alias for the star.  Columns 3 and 4
are the right ascension and declination of the star, in
degrees.  Column 5 is the EDR3 ID number.  Column 6 gives the 
$Gaia$ G magnitude.  Column 7 is the
environment class from K16 specifying the local environment
of the nebula as isolated (I; 155 instances), facing an 
\ion{H}{2} region (FH; 50 instances), within an \ion{H}{2}
region (H, 30 instances), or facing a bright-rimmed cloud
(FB, 30 instances).   An FB designation often entails an H
designation, so some objects have a compound environment
class (e.g., FB/H).  Columns 8 and 9 give the EDR3 parallax
and uncertainty, in milliarcsec. Columns 10--13 give the EDR3
proper motion in right ascension and declination, in
milliarcsec yr$^{-1}$, and their associated uncertainties. 
Column 14 is a code designating each star as part of the
Multiple Star Candidates group (1), Single Star Candidates  
group (2), or the Group 3 (3).  Column 15 is the spectral
type, either from Paper I where sources were classified into
three broad categories O/OB/B, or from the literature, in
which case the spectral type and luminosity class is
enclosed in parentheses.  Column 16 lists the literature
reference for the spectral classifications in
parentheses.   

The literature spectral types of
Table~\ref{tab:BSNkinematics} reinforce the conclusions of
Paper I that the central stars of SBNe are predominantly OB
spectral types. Of the 141 sources with spectral types, 136
are OB stars, three are M stars (possibly the evolved
descendants of massive stars), one is a carbon star, and one
is a K star.  As part of this work, we examined the  Wide
Field Infrared Explorer (WISE) and {\it Spitzer Space
Telescope}  archival images to re-evaluate local
environments of the SBN central stars using the criteria of
K16. We cannot determine that the arcuate ISM structures are
co-distant with the SBN central stars, only that these
structures are co-located on the sky. For the majority of
sources (65\%), we affirmed the environment class of K16. Of
the sources for which we reassign the local environment
class, 13 of 56 resulted from changing identifications
within complicated regions (e.g., updating a source from an
H to an FB where the rim of an interstellar bubble was
prominent at 8.0 $\mu$m). Nearly half, 24 of 56, involved
updating sources from I to FB or FH. These sources were
largely 5--15 arcmin from the associated \ion{H}{2} region
or bright-rimmed cloud,  many without sharp boundaries.  The
remaining 19 sources we reassigned the environment class
from I to H, where these stars and \ion{H}{2} regions appear
at the edge of WISE ATLAS tiles.  For the 50 sources in
Table~\ref{tab:BSNkinematics} drawn from the extended
bowshock candidate list of \citet{Jayasinghe2019}, the
listed environmental classes are new to this work.  

Table~\ref{tab:BSNkinematics2} lists kinematic measurements
for the SBN central stars. Column 1 is the identification
number, as in Table~\ref{tab:BSNkinematics}.  Columns 2 and
3 give the calculated \vtwod\ and associated uncertainty in
\kms. Where radial velocities were measured for the Multiple
Star Candidates and Single Star Candidates, Columns 4 and 5
provide the calculated \vthreed\ and associated uncertainty
in \kms. The largest source of uncertainty in the \vtwod\
calculations is the distance uncertainty arising from the
linear dependence of each  velocity component on distance.
The uncertainties on the angular  proper motions,
$\sigma_{\mu_\alpha}$ and $\sigma_{\mu_\delta}$, average
$<$1\%,  whereas the mean parallax uncertainty is 8\%,
thereby constituting the largest source of error on \vtwod.
Uncertainties on \vthreed\ will always be larger than the
uncertainties of \vtwod\ due to the additional uncertainty
term from the radial velocity measurement.  For the
multiple-star systems, this additional uncertainty may be
significant, $\geq$6 \kms\ as an observational minimum to as
much as 50 \kms\ for some binary systems where the systemic
radial velocity is poorly constrained from just a few
measurements. Column 6 lists the morphological position
angle ($PA_{\rm m}$) from the central star to the apex of
the SBN, in degrees from North toward East in Equatorial
coordinates, as tabulated in K16. This position angle was
measured by eye based upon {\it Spitzer Space Telescope} 24
$\mu$m  and WISE 22 $\mu$m archival images and carries an
estimated uncertainty of 8\degr.  Columns 7 \& 8 are the
kinematic position angle ($PA_{\rm k}$) of the star's
\vtwod\ vector and its associated uncertainty in degrees.
Columns 9 \& 10 give the difference between the
morphological position angle and the kinematic position
angle in degrees, \deltaPA=$PA_{\rm m}-PA_{\rm k}$, and
associated uncertainty. Column 11 provides the code, as in
Table~\ref{tab:BSNkinematics}, designating each source as
part of the Multiple Star Candidate group, the Single Star
Candidate group, or the Group 3 sources which lack radial
velocity measurements.   

\begin{figure}[ht!]
\fig{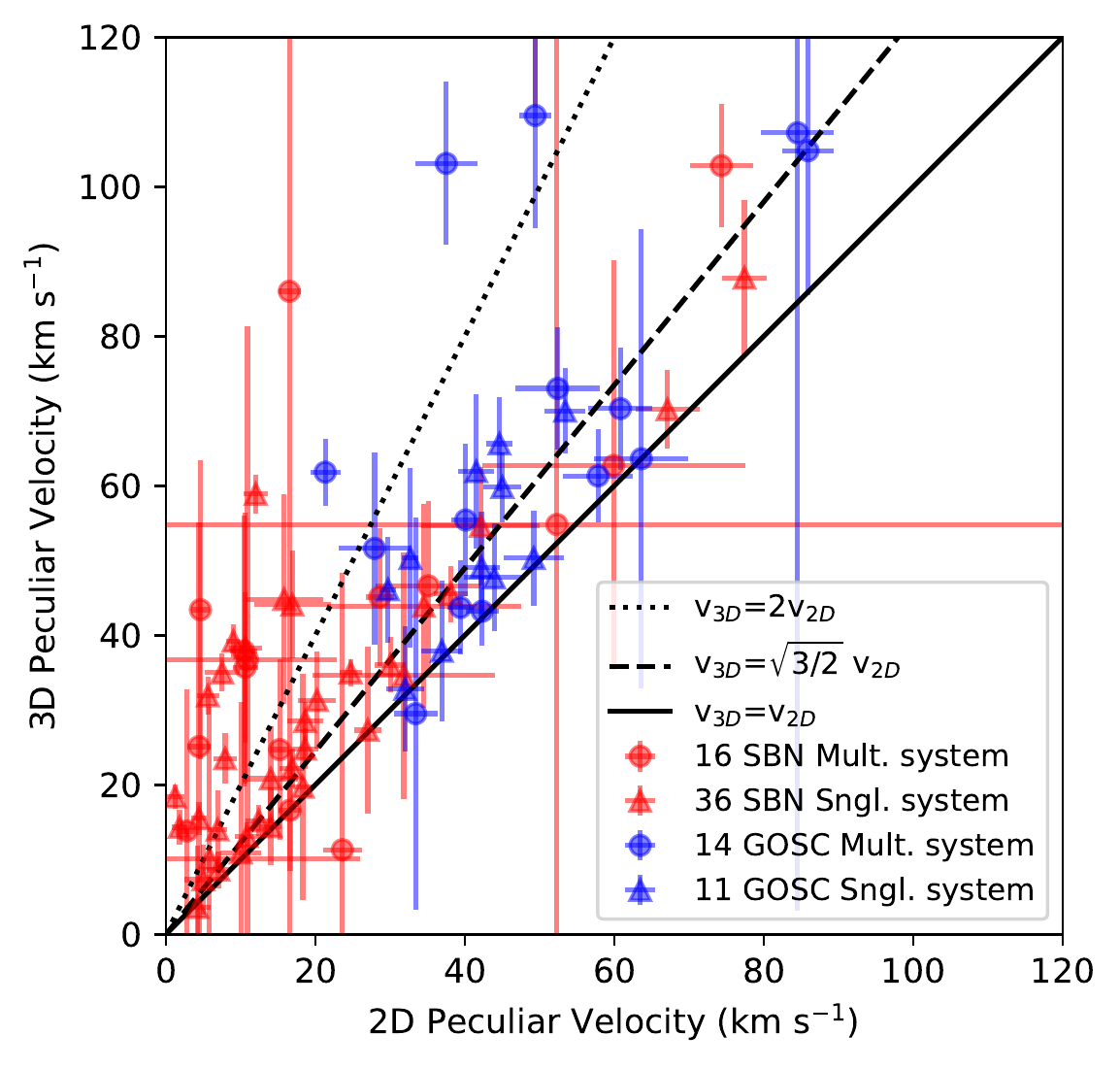}{4in}{}
\caption{Three-dimensional peculiar velocity versus two-dimensional
peculiar velocity for multiple star candidates (circles) and single-star candidates (triangles) 
from the SBN sample (red) and the GOSC comparison sample (blue). Only stars having multiple 
radial velocity measurements and meeting the astrometric selection criteria are shown.  A solid line marks the 1:1 correspondence, 
while a dashed line marks the $\sqrt{3/2}$:1 relation expected for isotropic velocities 
in three dimensions. 
\label{fig:twodthreed}}
\end{figure}

Figure~\ref{fig:twodthreed} plots the calculated \vthreed\
versus \vtwod\ for the \numspecGOSC\ systems in the GOSC
runaway sample ({\it blue symbols}) and the 52 SBN central
stars having radial velocity data ({\it red symbols}). 
Circles denote multi-star candidates, while triangles denote
single-star candidates. A solid black diagonal line marks
the 1:1 relation where \vthreed=\vtwod\ (i.e., where a
source would have zero peculiar radial velocity). A dashed
line marks the relationship for isotropic velocities where
the expected \vthreed = $\sqrt{3/2}$ \vtwod. A dotted line
indicates a \vthreed:\vtwod\ ratio of 2:1.  The majority of
SBN stars, 31 of 52 (60\%), lie between the solid and dotted
lines and are consistent within one sigma of the dashed
line,  indicating that their velocities are either entirely
within the plane of the sky or consistent with an isotropic
velocity distribution.  This agrees with the expectation
that morphologically selected SBNe central stars have
velocities primarily in the plane of the sky. If a star's
velocity were primarily radial through a stationary ISM, the
morphology of the nebula would appear more circular than
arcuate \citep{Meyer2016}. For a minority of sources (40\%),
particularly in the low-\vtwod\ regime ($\lesssim$10 \kms),
the ratio \vthreed:\vtwod\ is greater than 2. These are
likely to be  either unidentified binary systems or
multiple-star systems with a poorly determined systemic
velocity, a consequence of having only a few radial velocity
measurements. One system, the SBN multiple-star candidate
346 with \vthreed=220 \kms, lies off the top left of the
plot. Upon closer examination, SBN 346 appears less arcuate
and less symmetric than the other sources in the K16
catalog. Given its large peculiar radial velocity of
196$\pm$9 \kms\ and dubious nebular morphology, we
reclassify this as a doubtful SBN candidate.  It may be a
radial velocity runaway or a multiple-star system with a
poorly determined radial velocity.

For some SBN central stars the calculated \vtwod\ may not be
representative of the true star-ISM velocity differential if
there exists a significant local flow of interstellar
material.  Such a local  flow will be a significant factor
in determining the  nebular morphology and the relative
star-ISM velocity, most dramatically so if the  star's
peculiar velocity is small. Winds from \ion{H}{2} regions
may impart local flow velocities   as large as 30 \kms\
\citep{TenorioTagle1979, Bodenheimer1979, Castaneda1988}. 
If a local flow is transverse, the star-ISM velocity
differential would also be transverse creating the arcuate
morphology without need for a stellar peculiar velocity. 
This may help explain why stars with peculiar velocities
seemingly dominated by radial velocity components exhibit
arcuate SBN.  This class of objects would include the
``in-situ'' bowshock candidates where an arcuate nebula
points toward a nearby \ion{H}{2} region (e.g., our FH
environment class).  Of the \numBSN\ SBN objects in this
sample, 50 (19\%) have FH environmental classifications,
meaning that a non-negligible fraction could be influenced
by local ISM flows.  

\subsection{The Runaway Fraction of SBN Central Stars}

\begin{figure}[ht!] 
\fig{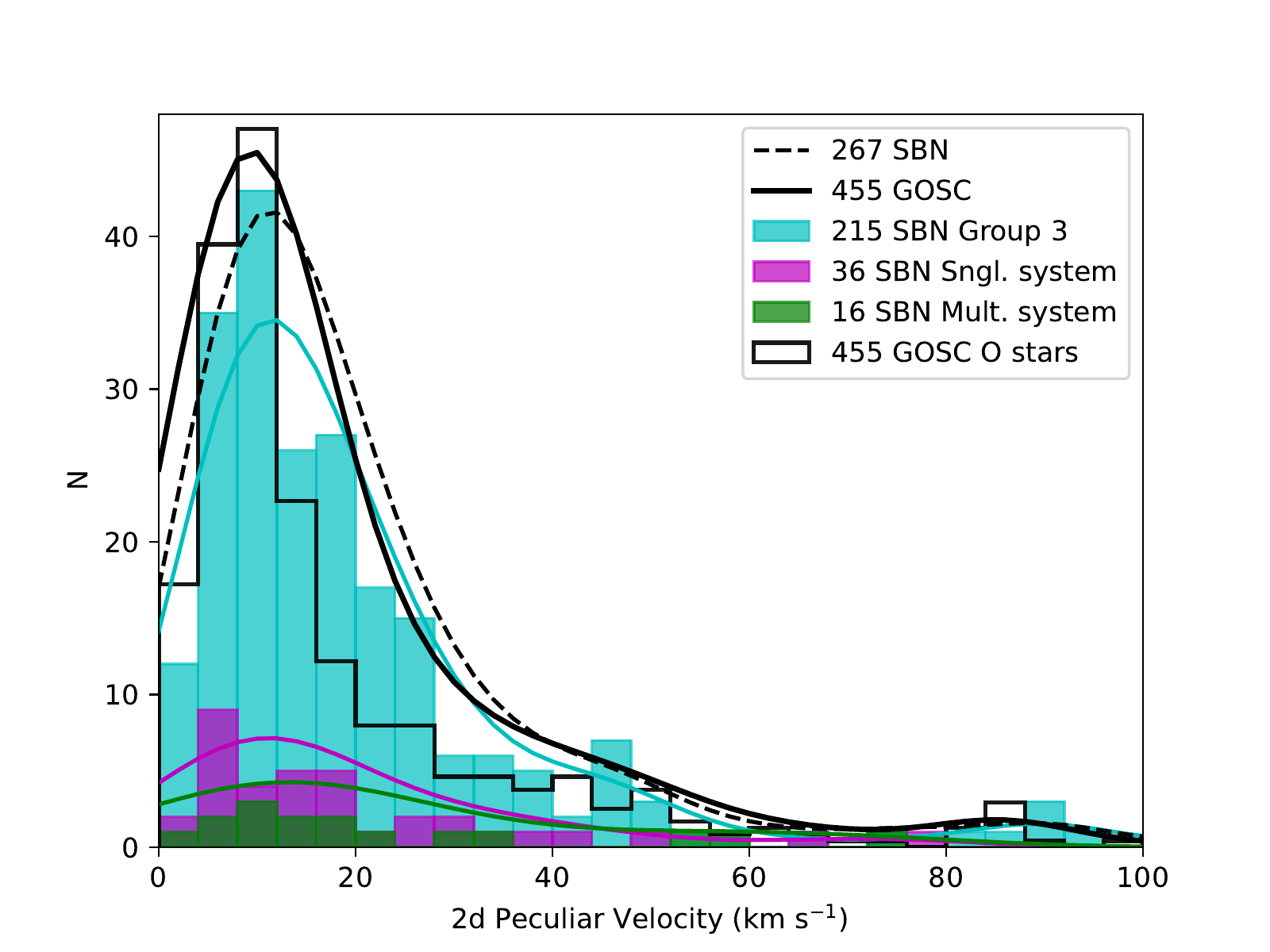}{4in}{}
\caption{Histogram of \vtwod\ peculiar velocities for
subsamples  of SBN  central stars: the Multiple-Star
Candidates ({\it green}), the  Single-Star Candidates ({\it magenta}),
and the Group 3 sample ({\it cyan}).   Curves denote Gaussian
probability density estimates for each subsample of SBN
stars   and for the  entire sample of \numBSN\ SBN central
stars ({\it black dashed curve})  and the GOSC sample ({\it black solid 
curve}). The distributions of the SBN sample and the GOSC O-star 
sample are statistically different. \label{fig:histPDF}}
\end{figure}

Figure~\ref{fig:histPDF} presents histograms and kernel
density estimates (KDEs) for the three SBN central star
groups: Group 3 systems (cyan), the Single-Star Candidates
(magenta), and the Multiple-Star Candidates (green). Colored
curves  denote the KDE for each group and the full SBN
sample ({\it black dashed curve}).  An unfilled black
histogram and  black solid curve designate the distribution
and KDE of the \numGOSC\ GOSC stars, renormalized to
facilitate comparison.    The KDEs for each population show
a distinct peak centered near 9 \kms\ and a long tail out to
80 \kms\ and beyond. Although the KDEs suggest the Single
Star Candidates may contain a high fraction of runaway
stars, this may be a result of the small number of stars in
that group. All SBN groups are consistent with having been
drawn from the same parent population with the p-values
(from K-S tests) $>$0.30 for all sample comparisons. 
However the SBN and the GOSC distributions are different at
the $p$=0.003 level.  This difference is mostly driven by
the B stars within the SBN sample, as a comparison of just
the SBN O stars with the GOSC sample results in  $p$=0.042,
marginally consistent with being drawn from the same
population.   The SBN sample is noticeably shifted toward
larger velocities and has fewer objects at small
velocities.  The 16th/50th/84th percentile velocities of the
SBN sample are 6.5/14.6/32.6  \kms\ versus 5.6/11.4/32.6
\kms\ for the GOSC sample. This difference could be
attributed to the larger distances for the SBN stars (mean
distance of 2300 pc) than the GOSC stars (mean distance of
3300 pc), since transverse velocity inferred from proper
motion scales linearly with distance.

From the \vtwod\ velocities of the \numBSN\ SBN central
stars we identified 63$^{+25}_{-18}$  candidate runaways with
\vtwod$>$25 \kms.  This implies a runaway fraction of 
24$^{+9}_{-7}$\%, nearly identical to that of the GOSC O stars.  

Among the 142 stars with spectral classifications from the
literature or our work (Paper I),  43 (30\%) are O stars, 61
(43\%) are early B stars, with the remainder (26\%)  being
indeterminate OB stellar types on the O9/B0 boundary  (other
than the six late-type stars F--M).  Of the 43 O stars,
12$^{+2}_{-1}$ have \vtwod$>$ 25 \kms\ (28$^{+4}_{-2}$\%).  
This O-star runaway fraction is larger than the 22$\pm$3\%
observed among the GOSC stars and is consistent with the
high end of the observed O star runaway fractions
\citep[e.g.,][]{Blaauw1961,Stone1991,Oh2015}.  Among the 61
B stars, 16$^{+4}_{-1}$  (26$^{+7}_{-2}$\%)  are runaway
star candidates, much larger than the 2\% B star runaway
fractions measured by \citet{Blaauw1961} and
\citet{Gies1986}.   For the sample SBN central stars, it
appears that O and B stars have similar runaway fractions.
 Among the 16 Multiple-Star Candidates  3$^{+2}_{-1}$
(18$^{+18}_{-12}$\%) has \vtwod$>$25 \kms , whereas
9$^{+2}_{-2}$ (25$^{+5}_{-6}$\%) of the 36 Single Star
Candidates are runaway stars. Within the Poisson
uncertainties, multiple-star systems and candidate
single-star systems have similar runaway fractions.

\subsection{Morphological and Kinematic Position Angles}

\begin{figure}[ht!]
\fig{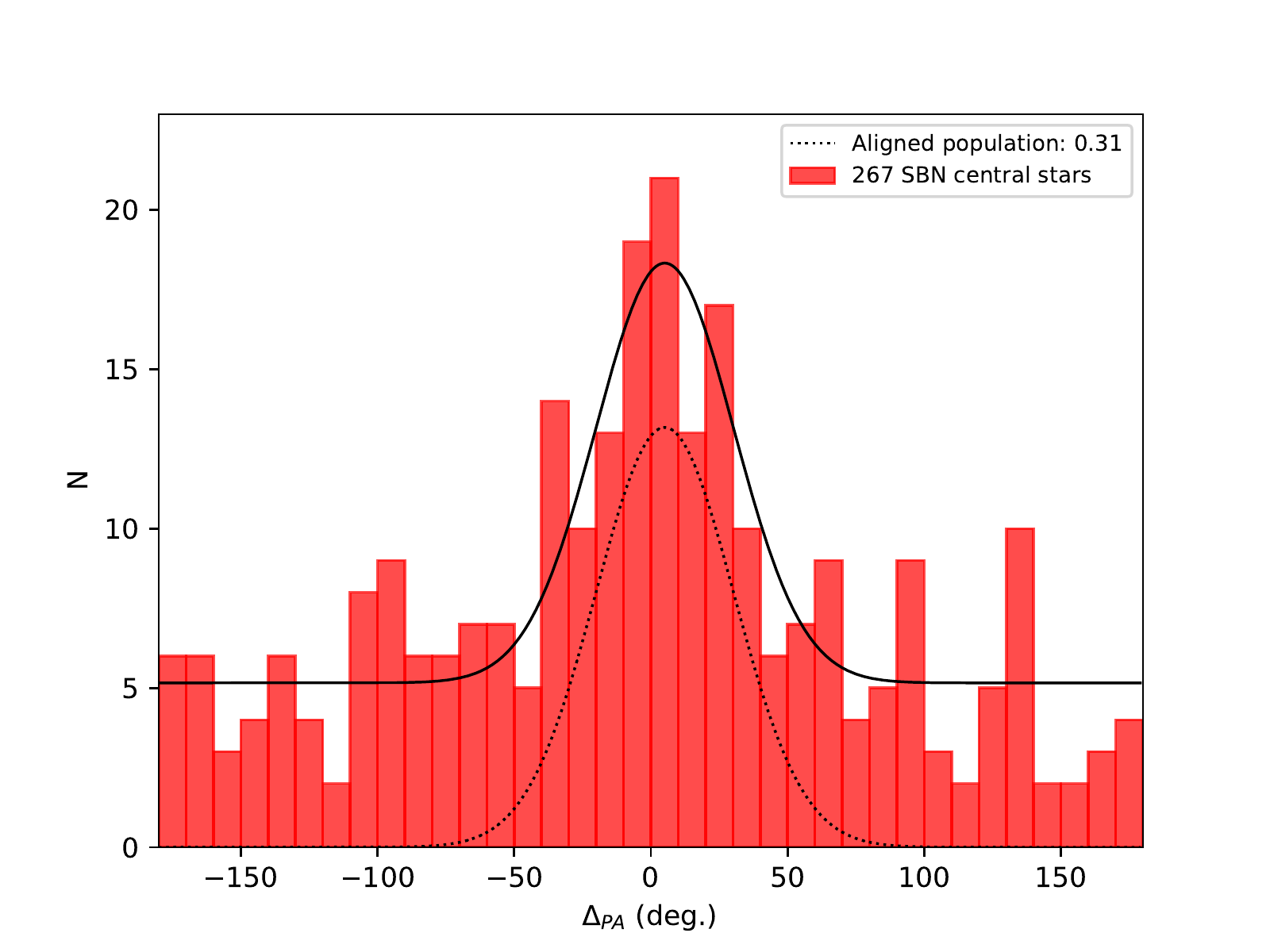}{5in}{}
\caption{Histogram of $\Delta_{PA}$, the difference between the 
morphological position angle of the arcuate nebulae
and the position angle of the central star's projected peculiar velocities. 
The distribution is well approximated by the sum ({\it solid curve}) of 
a Gaussian component representing a highly aligned population ({\it dashed curve})
and a random (flat, non-aligned) component.   
\label{fig:histDeltaPA}}
\end{figure}

We used the computed \deltaPA\ parameter---the difference
between morphological and kinematic position angles---to
assess the degree of  alignment between the SBN
morphological position angles and the stars' peculiar
velocity vectors. Figure~\ref{fig:histDeltaPA} presents a
histogram of the \deltaPA\ distribution for the \numBSN\ SBN
central stars, using bin widths of 10\degr. The Figure shows
a peak near \deltaPA=0\degr, indicating strong alignment
between the morphological axes of the SBN and the peculiar
motion vectors.  This peak is present regardless of the
choice of Solar peculiar motions (see Appendix), but is more
pronounced by about 50\% when the adopted [$U_\odot$,
$V_\odot$, $W_\odot$] values are used.  There is a broad
tail to either side, with some sources showing
anti-alignment (\deltaPA=$\pm$180 \degr).  The distribution
can be approximated by two hypothetical populations: a
highly aligned population centered near 0\degr\ having a
Gaussian width of $\sigma$=25\degr\ (dashed curve) and a
random (non-aligned) population.   The solid black curve
depicts the sum of these two components.   The ``highly
aligned population'' comprises 31\% of the sample, while
the  ``random population'' comprises 69\%.   The overall
distribution is similar to that presented in K16 (Figure 12)
but contains more sources, uses better astrometric data, and
corrects for solar motion relative to the Local Standard of
Rest.  

The two populations identified on the basis of
Figure~\ref{fig:histDeltaPA} suggest that, in a minority of
cases, the stellar peculiar velocities are responsible for
shaping the orientation of the nebulae.  In a majority of
cases, there is considerable dispersion between the star's
velocity vector and the morphological axis of the nebulae. 
This suggests that local environmental factors, such as
local ISM flows or the proximity to an \ion{H}{2} region or
bright-rimmed cloud undergoing photoevaporation---rather
than the stellar motion---dominates the formation of the
nebulae in most instances.  Notably, in about 12\% of the
sources, $\vert$\deltaPA$\vert$$>$100\degr,  indicating
anti-alignment between morphological and kinematic position
angles.  Such anti-alignment could occur in cases where an
OB star is ejected form a nearby \ion{H}{2} region and  an
outflow from that \ion{H}{2} region impinges upon the
star.   

\begin{figure}[ht!]
\fig{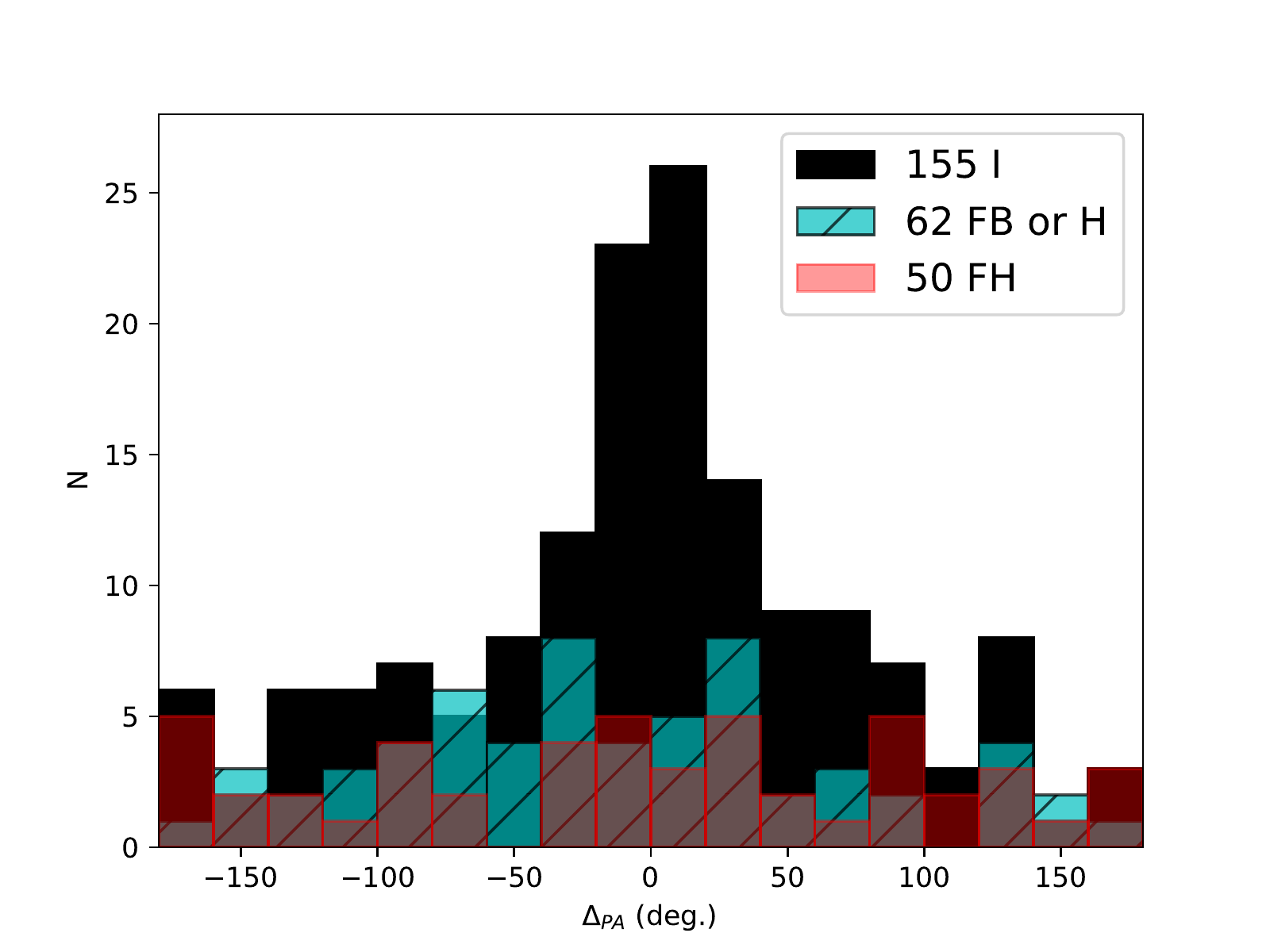}{5in}{}
\caption{Histogram of $\Delta_{PA}$, with systems color coded by 
local environment class. Isolated systems peak strongly near zero,
indicating alignment between morphological and kinematic vectors.  
FB/H and FH stars have a nearly flat distribution.   
\label{fig:histbyEnv}}
\end{figure}

Figure~\ref{fig:histbyEnv} plots a histogram of \deltaPA\
color coded  by environment class, with solid black bars
denoting stars in isolated environments (I), blue hatched
bars denoting stars either inside \ion{H}{2} regions (H) or
facing bright-rimmed clouds (FB), and red transparent bars
depicting stars facing toward nearby \ion{H}{2} regions.
Bin  widths are twice as large as in
Figure~\ref{fig:histDeltaPA} owing to the smaller numbers of
sources in the latter two groups. We merge the H and FB
designations, as stars inside \ion{H}{2} regions often face
bright rims lining those cavities and may encounter similar
photoevaporative flows. The largest group, the isolated
stars (I), shows a similar distribution of \deltaPA\ to the
general population in Figure~\ref{fig:histDeltaPA}.  Of the
155 stars in isolated environments, 115 (74\%) are highly
aligned, having $\vert$\deltaPA$\vert$$<$45 \degr.  The two
remaining groups, H/FB and FH, both have flatter
distributions with small, but not statistically significant
overdensities near \deltaPA = 0\degr.  Assessed another way,
the Pearson correlation coefficient between $PA\_m$ and
$PA\_k$ is $r$=0.37 ($p$=10$^{-6}$) for the 155 environment
class I stars, confirming the strong correlation.   The 62
FB/H environment class stars have  $r$=0.11 and $p$=0.36,
indicating  a weak or no significant correlation.  The 50
environment class FH stars also show no correlation
($r$=0.04, $p$=0.75).  The weak or non-existent correlation
for FB/H stars is unsurprising,  as we would expect
significant small-scale flows in the proximity of bright
rimmed clouds and  bubbles where photoevaporative flows are
common and significant relative to the peculiar velocity of
the stars.   The lack of correlation for FH stars is also
unsurprising.  We would also expect flows emanating from
\ion{H}{2} regions---at velocities that may dominate the
relative  star-ISM differential in cases where the star's
velocity is small.  Finally, we stress that our
environmental classifications are an imprecise
characterization of the stars' true local conditions, based
only on  angular separations from prominent infrared
features.  A more quantitative  analysis comparing SBNe
three-dimensional Galactic positions with those of OB stars
and \ion{H}{2} regions would improve the reliability of the
environmental  classifications. Such an effort should now be
possible but is beyond the scope of this work.   

\begin{figure}[ht!]
\fig{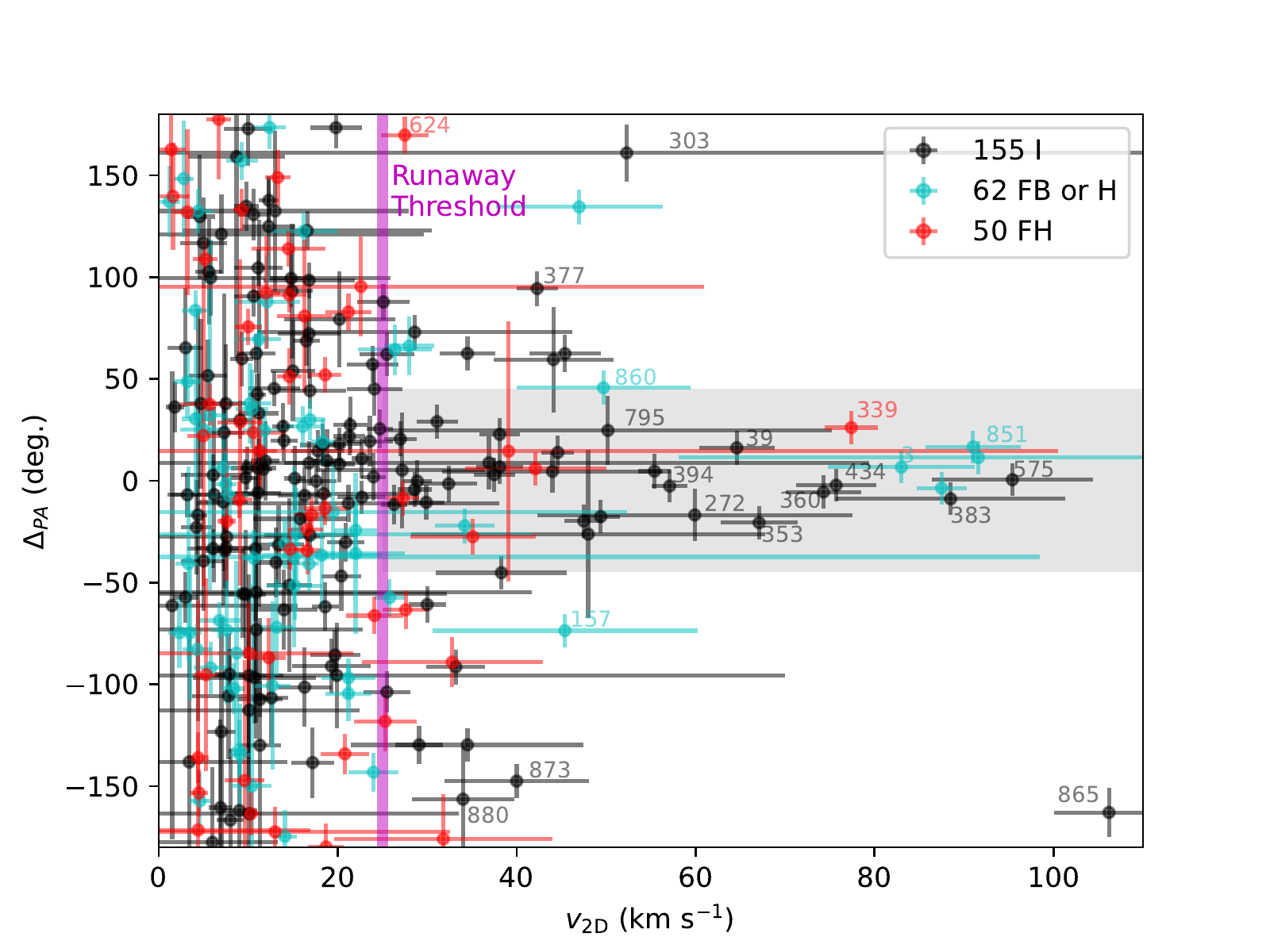}{5in}{}
\caption{$\Delta_{PA}$ versus \vtwod\ peculiar velocity. Colors distinguish 
the I, FB/H, FH environment classes. The gray shaded region marks the regime of 
runaway stars (\vtwod$>$25 \kms) that have nebular morphological position angles 
  aligned with the central stars' velocity vectors 
  ($\vert$\deltaPA$\vert<$45\degr).
\label{fig:deltaPAversusVel}}
\end{figure}

Figure~\ref{fig:deltaPAversusVel} plots \deltaPA\ versus \vtwod\
for the SBN sample. Black symbols and error bars mark 
isolated stars (I), cyan symbols mark stars inside \ion{H}{2} and
facing bright-rimmed clouds (H/FB), and red symbols mark stars with SBN
facing nearby \ion{H}{2} regions.  Numerical labels identify
some of the highest velocity stars.  A vertical magenta bar
delineates the runaway threshold of 25 \kms.  The gray shaded
rectangular region marks the regime where morphological
positions angles are highly aligned with the kinematic
vectors  ($\vert$\deltaPA$\vert$$<$45\degr) and \vtwod$>$25
\kms.  With few exceptions, the stars with the largest
velocities also have the smallest \deltaPA, falling within
the shaded region.   Forty-four of the 155 isolated stars
(28$\pm$4\%), eight of the 62  FB/H stars (13$\pm$5\%), and
ten of the 50  FH stars  (20$\pm$6\%) are runaways. 
Thirty-eight of the 63 runaway stars (80\%) show strong
alignment (within 45\degr) between their morphological and
kinematic position angles.  Among stars with projected
velocities $<$25 \kms, there is a large dispersion in
\deltaPA.   Evidently factors other than peculiar velocity 
determine \deltaPA\ for
these nebulae, such as local ISM flows that are on the order
of the $\approx$10 \kms\ peculiar velocities measured for
this population.   Other factors may include density
inhomogeneities in the vicinities of  spiral arms, molecular
clouds, or star clusters that create departures  from the
smooth circular Galactic rotation curve assumed in computing
\vtwod. 

Among the runaways a few stars show dramatic  misalignment,
having \deltaPA$>$120\degr.  Star 303 (labeled in
Figure~\ref{fig:deltaPAversusVel}) is among the most distant
in the sample ($d$$>$10 kpc at Galactic longitude
$\ell$=60\degr), giving it a parallax uncertainty that
results in a very large uncertainty on \vtwod\
($\sigma_{v2D}=68$ \kms).  Formally, it is consistent with
having a negligible (non-runaway) peculiar velocity. Star
865 (\vtwod=108 \kms)  is also quite distant ($>$ 5 kpc at
$\ell$=340\degr) but has well-measured kinematics.  It
appears to be confined within an interstellar cloud or
bubble irradiated by an external source of illumination, so
the tabulated morphological position angle probably
corresponds to the direction of this radiation and not the
stellar velocity vector.  The nebulae preceding Star 873 is
compact and appears isolated from external influences; we
have no good explanation for the large
\deltaPA=$-$147\degr.  Star 880 has a well-defined nebula
and well-measured kinematics, yet its morphological
orientation is nearly 180\degr\ from its projected velocity.
Star 624 faces an \ion{H}{2} region, has a velocity just
above the 25 \kms\  threshold, and shows anti-alignment at
\deltaPA=170\degr.

\clearpage

\section{Conclusions}

The Gaia EDR3 astrometric data and new multi-epoch optical
spectroscopy has allowed us to investigate the kinematic
properties of a large sample of morphologically selected
stellar bowshock nebulae and their central stars. 
Quantities of particular interest for addressing the origins
of bowshock nebulae and  runaway stars are the stellar
peculiar velocities (both 2D and 3D), their multiplicity, as
evidenced by the dispersion in radial velocity measurements,
and the degree of angular alignment between the SBNe
morphological axes and stellar kinematic vectors.   For
comparison we have computed the same quantities for a
control sample of Galactic O stars.  Table~\ref{tab:summary}
provides a summary of key statistics derived from this study
regarding runaway fractions, multiplicity, and alignment of
SBN morphological and central star kinematic axes.

Based on the \vtwod\ velocities the runaway fraction for SBN
central stars is 24$^{+9}_{-7}$\%, nearly identical to the
22$^{+3}_{-3}$\% runaway fraction for GOSC stars. The SBN
runaway fraction is  consistent with the high end of
simulations \citep{Oh2015} and observations
\citep{Blaauw1961,Stone1991},  but generally larger than
most estimates.  The SBN sample consists primarily of early
B stars, and so the runaway fractions are much larger than
other estimates  of B stars which lie near 2\% \citep[e.g.,
][]{Blaauw1961, Gies1986}.  We conclude that while the
runaway rate of SBN central stars is likely enhanced
relative to the general OB population, runaway stars do not
comprise a majority fraction of SBN central stars.  The
median \vtwod\ of our sample is 14.6 \kms, far below runaway
thresholds and contrary to some common assumptions that 
bowshock central stars have large peculiar velocities.  
Both the Multiple-Star Candidates group and the Single-Star
Candidate group show similar  runaway fractions, indicating
that the mechanism responsible for their large velocities
apparently acts on binary systems with equal efficiency as
single stars.  This observation would seem to favor the DES
(N-body interactions in clusters) over the BSS (supernova
ejection)  scenario.   

Four of the 13 SBN runaways that have multiple spectroscopic
measurements show evidence for binarity,  yielding a minimum
multiplicity fraction for runaways of 30$^{+16}_{-15}$\%. 
Among the GOSC runaways in Table~1,  this minimum
multiplicity fraction is $>$28\% after corrections for false
positives arising from potential line profile variations
caused by atmospheric pulsation.  Both of these fractions
are much larger than the $\sim$few percent proposed by most
pure BSS ejection simulations \citep[e.g.,][]{Eldridge2011,
Renzo2019}. DES ejection simulations predict 5--20\% 
\citep[e.g.,][]{Oh2015, Perets2012}, in better agreement
with---but still markedly lower than---the observed
fractions for SBN central stars and Galactic O stars. If SBN
central stars are representative of the population of field
OB stars, as is suggested by the consonance between the
samples' multiplicity fractions and runaway fractions, this
suggests that the DES contributes more heavily to the field
OB population than BSS.   The significant numbers of
multiple systems ejected from young stellar clusters
\citep{Maiz2022} adds evidence to this scenario.
Alternatively, given the high fraction of SBN central stars
in isolated environments with low peculiar  velocities
($\simeq$14 \kms), this may provide evidence that a fraction
of massive stars may form in relative isolation. However,
without attempts to trace these stars back to known star
forming regions and OB associations, this result is largely
conjecture and begs future follow-up studies that attempt to
link  isolated massive stars to their birth locations. 

About one-third (31\%) of SBNe have morphologies that are
well-aligned with the motion of their central stars---a
``highly aligned population''.   The majority (69\%) show
weak or no  alignment---``the random population''.  The
degree of alignment increases with peculiar velocity
(Figure~\ref{fig:deltaPAversusVel}). The relative lack of
alignment for low-velocity stars implicates the effects of
local ISM flows  particularly in the vicinity of energetic
\ion{H}{2} regions.  Indeed, the sources showing low degrees
of alignment tend to be those facing or within \ion{H}{2}
regions or facing bright-rimmed clouds where local flows are
expected to play a role (Figure~\ref{fig:histbyEnv}).  
However, some seemingly isolated nebulae also show low
levels of alignment.  Given the low stellar peculiar
velocities (median of 14.6 \kms\ for SBN stars), these flows
do not need to be especially fast to influence the
orientation of the nebula.   The flattened distribution of
\deltaPA\ for weakly aligned sources
(Figure~\ref{fig:histDeltaPA}),  may indicate that the
majority of SBN central stars are isotropically distributed
``walkaways''  acting as the ``interstellar wind vanes''
\citep{Povich2008}.   Enhancements in gravitational
potential, such as along spiral arms, may also play a role
in creating local deviations from the smooth circular
Galactic rotation curve assumed in our analysis, at the
level of 5--10 \kms.   Given the consistency of SBN central
star properties to those of O stars in the field, stellar
bowshock nebulae are a useful infrared indicator of the
presence of an OB star, particularly for identifying field
populations that lie far from young clusters and OB
associations.

\acknowledgments

An early version of this manuscript appeared  as Chapter 3
in the PhD dissertation of W.T. Chick \citep{Chick2020b}. We
thank Jes\'us Ma\'iz Apell\'aniz and Doug Gies for helpful
comments on an early version of this manuscript. 
We thank students Logan Jensen, Harrison S. Leiendecker,
Jacob N. McLane, Evan Haze Nunez, Jason A. Rothenberger, and
Ashley N. Piccone for taking observations that contributed
to this work. Our team is grateful for support from the
National Science Foundation through grant AST-1412845,
AST-1411851, AST-2108347, and REU grant AST-1063146, as well as NASA
through grant NNX14AR35A. This work has made use of data
from the European Space Agency (ESA) mission Gaia
(https://www. cosmos.esa.int/Gaia), processed by the Gaia
Data Processing and Analysis Consortium (DPAC,
https://www.cosmos.esa.int/web/Gaia/dpac/consortium).
Funding for the DPAC has been provided by national
institutions, in particular the institutions participating
in the Gaia Multilateral Agreement. 

\vspace{5mm}

\facilities{WIRO, APO}

\software{IRAF \citep{Tody1986}, astropy \citep{Astropy2013}, SciPy \citep{Virtanen2020}}

\clearpage

\appendix

\section{A New Approach to Computing Local Standard of Rest Velocity}

Astrometric and radial velocity measurements in the
Heliocentric frame of reference require  corrections for
Solar motion relative to the Local Standard of Rest before
conversion to a rotating Galactic reference frame.  Many
works have calculated [$U_\odot$, $V_\odot$,
$W_\odot$], the solar motion relative to the LSR toward
Galactic center, toward the direction of prograde Galactic
rotation, and toward Galctic North, respectively. A wide
range of values has been reported---$U_\odot$=6--12 \kms,
$V_\odot$=3--26 \kms, $W_\odot$=5--9 \kms---even as improved
kinematic data, larger stellar samples, and better
understandings of systematic biases by stellar age have
emerged \citep[e.g., ][]{Mayor1974, Dehnen1998, Bobylev2007,
Aumer2009, Schonrich2010, Golubov2013, Ding2019}.  The
sample under consideration here, OB stars, represent a young
stellar population that likely suffer from some of the same biases
(i.e., a low velocity dispersion relative to older
populations)  as discussed in these works.  Our attempts to
adopt  recent estimates for  [$U_\odot$, $V_\odot$,
$W_\odot$] \citep[e.g., ][]{Schonrich2010, Ding2019}
yielded  stellar kinematic vectors that showed physically implausible trends 
by Galactic longitude.  This led us to
pursue using the SBN sample to derive a new estimate of the
Solar motion parameters.  Our intent was not a repudiation of
other recent measurements of [$U_\odot$, $V_\odot$,
$W_\odot$], which are based on more extensive analyses of
larger stellar samples, but rather an {\it ad hoc}
measurement based on a fundamentally new approach  that
yields physically plausible results for the SBN OB
stars.  Our results turn out to be consistent with the range
of other recent Solar motion velocities.

Our initial analyses using either the \citet{Schonrich2010}
[$U_\odot$, $V_\odot$, $W_\odot$]= [11.1, 12.2, 7.2] \kms\ 
or the \citet{Ding2019} [U$_\odot$, V$_\odot$, W$_\odot$]=
[8.63, 4.76, 7.26] \kms\ solar velocities produced
systematic trends in stellar kinematic position angle with
Galactic longitude.   Using the former set of Solar
velocities yielded SBN central star velocities that were
systematically  directed toward smaller longitudes in the
first quadrant and larger longitudes in the fourth
quadrant---a result that is physically implausible.  Under
the latter assumption  the opposite systematic trend
obtained---also implausible.  An incorrect Solar motion
vector can explain these systematic effects. 

The choice of stellar population used to infer [$U_\odot$,
$V_\odot$, $W_\odot$]  changes the computed values, most
dramatically so with $V_\odot$ \citep[e.g., see discussion
in][]{Aumer2009, Golubov2013}.   Young populations (OB
stars) have colder kinemtics (smaller velocity dispersion) than older
populations which have had time to be heated within the
Galactic potential.  For the analyses in this work we chose
to estimate an {\it ad hoc}  solar motion using our SBN
dataset itself by applying a set plausibility constraints
that serve as boundary conditions for our estimation.

\begin{enumerate} 

\item{The kinematic vectors of SBN central
stars should not correlate with Galactic longitude. 
Although our sample lies mostly in the first and fourth
quadrants, their peculiar motions should have no preferred
orientation by quadrant in order to be consistent with the expected isotropic velocities. } 
\item{The dispersion between the nebular morphological position 
angles and the central star's kinematic position angles should be minimized. Although one of 
the purposes of this work is to explore the extent to which morphological and kinematic position angles align, 
such an alignment is already apparent even without correction for LSR Solar motion.  Taken as a whole,
the sample distribution of SBN nebular position angles should be maximally consistent with
the distribution of central star kinematic position angles.   }
\end{enumerate}

\begin{figure}[ht!]
\fig{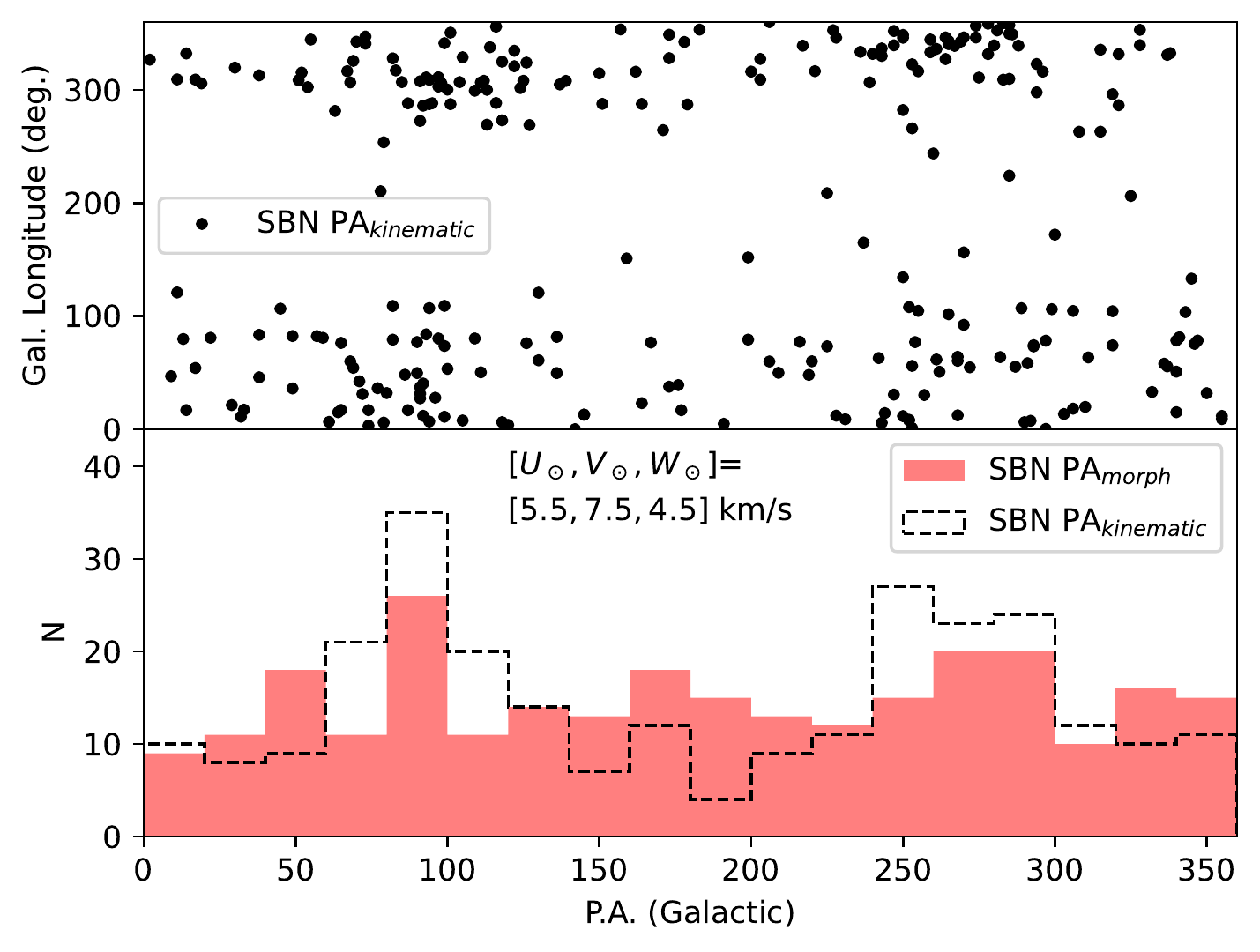}{5in}{}
\caption{({\it upper panel}) Galactic longitude versus kinematic position
angle for the SBN central star sample.  ({\it lower panel}) Distributions of SBN 
morphological position angles ({\it red histogram}) and central star kinematic position 
angles ({\it dashed histogram}). The two
are consistent with having been drawn from the same parent population. 
\label{fig:A1}}
\end{figure}

Figure~\ref{fig:A1} illustrates the distributions of
kinematic position angles  for SBN central stars resulting
from Solar motion parameters [U$_\odot$, V$_\odot$,
W$_\odot$]= [5.5, 7.5, 4.5] \kms, adopted as values that best satisfy the above 
bounday conditions.  The upper panel plots 
Galactic longitude versus  kinematic position angle in
Galactic coordinates.  Here, it is clear that the sample
preferentially lies in the first and fourth Galactic
quadrants with few objects in the second and third
quadrants.  There is no probable correlation ($p$=0.36) 
between longitude and position angle, as expected for
isotropic stellar motions. The lower panel shows histograms
of the bowshock nebular morphological position angles in
Galactic coordinates ({\it red filled histogram}) and the
central star kinematic position angles ({\it dashed black
histogram}). A Kolmogorov-Smirnov (K-S) test shows that the
two distributions have a 52\% probability of being drawn
from the same parent distribution.  Both histograms show
excesses near Galactic position angles 90\degr\ and
270\degr, i.e., along the Plane.  A similar excess was
noted  for the SBN parent sample morphological position
angles in \citet{Kobulnicky2016}.  We understand this excess
to be a selection effect rather than a violation of
isotropy.  High-velocity OB stars  having vectors orthogonal
to the Plane will traverse the $\approx$80~pc scale height
\citep{Clemens1988} of the inner disk's molecular layer even
in their short lifetimes, making them less likely to
generate detectable bowshock nebulae.  Coversely, OB stars
having velocity vectors parallel to the Plane will remain
within this higher density layer longer and produce observable
bowshock nebulae for a greater fraction of their
lifetimes. 

Our adopted [U$_\odot$, V$_\odot$, W$_\odot$] not only 
provide stellar motions that pass the self-consistency
checks outlined above, they also lie within the range of
values obtained by other (more traditional) types of
analysis, despite being grounded in a fundamentally new
methodology. U$_\odot$=5.5 \kms\ lies at the low end
of---but consistent with---published ranges \citep[see Table
1 of][5.5--11.7 \kms]{Ding2019}.  V$_\odot$=7.5 \kms\ falls
near the median of published values (3--14 \kms, with a
couple estimates near 20 \kms).  Our W$_\odot$=4.5 \kms\ is
considerably smaller than the narrow range of 6.5--7.5 \kms\
generally produced by other studies.  Values larger than 5.0
\kms\ produce significantly poorer agreement with our stated
boundary conditions.  It may be the case that  the OB
central stars of SBN have a systematically smaller $z$
component relative to other (older) stellar populations that
inform other studies.   We do not attempt to estimate
uncertainties on [U$_\odot$, V$_\odot$, W$_\odot$] owing to
the small number of objects involved and the imprecise
nature of the boundary conditions adopted.  Uncertainties of
$\pm$1.5 \kms\ are plausible and used in Monte Carlo error
propogation, where needed.

\clearpage

\movetabledown=1.9in
\begin{deluxetable*}{lrrrrrrrrcc}
\tabletypesize{\scriptsize}
\tablecaption{Kinematic Data of 25 Runaway GOSC Stars \label{tab:GOSCRunaway}}
\tablewidth{0pt}
\tablehead{
\colhead{GOSC ID}&\colhead{Alias} &\colhead{$RUWE$} &\colhead{\vtwod}  & \colhead{$\sigma_{\rm v2D}$} & \colhead{$\overline{RV}_\odot$}&\colhead{$\overline{\sigma_{RV}}$} &\colhead{RV RMS} & \colhead{} $P$($\chi^2$,$\nu$) & \colhead{SpecType} & \colhead{Obs. Dates} \\
\colhead{}       &\colhead{}      &\colhead{}       & \colhead{(\kms)} & \colhead{(\kms)}         & \colhead{(\kms)}	     & \colhead{(\kms)} 	      &\colhead{(\kms)} & \colhead{}	& \colhead{}		 &\colhead{(HJD$-$2.458,000)}	 \\   
\colhead{(1)} &\colhead{(2)}   & \colhead{(3)}      & \colhead{(4)}    & \colhead{(5)}            & \colhead{(6)}	     & \colhead{(7)}   & \colhead{(8)} & \colhead{(9)} & \colhead{(10)} & \colhead{(11)}
}
\startdata
20   & ALS 11244   & 1.047&  30  &   1   &  -39 & 18 &   7 & 0.981 & O5If  &  628.96, 674.98, 712.97, 747.77,786.55  \\
27   & ALS 12688   & 1.034&  85  &   5   & -184 & 30 & 104 & 0.000 & O5.5V &  668.97, 694.84, 747.79, 779.56, 786.56 \\  
78   & BD+39 1328  & 1.298&  38  &   4   & -112 &  5 &  10 & 0.004 & O8.5Iab &  764.96, 724.97, 731.97 \\                
80   & BD+43 3654  & 0.937&  45  &   2   &  -53 & 10 &   6 & 0.798 & O4If &  628.96, 712.97, 747.78, 779.56, 786.56 \\   
82   & BD+50 886   & 1.043&  37  &   3   &  -36 &  9 &   9 & 0.524 & O4V &  668.96, 724.95, 731.91 \\
88   & BD+60 134   & 0.930&  86  &   3   & -100 & 11 &  19 & 0.027 & O5.5V &  712.98, 724.95,725.92, 779.58, 786.57 \\   
90   & BD+60 261   & 0.882&  45  &   2   &  -77 &  4 &   4 & 0.405 & O7.5III &  712.99, 725.94, 779.59, 786.57 \\
106  & BD-10 4682  & 1.096&  33  &   3   &    6 & 14 &  26 & 0.031 & O7Vn &  747.57, 752.57, 779.54 \\
\enddata
\tablecomments{(1) GOSC source identification number, (2) common alias, (3) Gaia EDR3 re-normalized unit weight error, 
(4) projected two-dimensional peculiar velocity \vtwod, (5) uncertainty on \vtwod, (6) mean Heliocentric radial velocity, 
(7) mean radial velocity uncertainty, (8) radial velocity RMS, (9) probability of the null hypothesis; 
probabilities $<$0.05 are considered multiple-star candidates, 
(10) literature spectal type and luminosity class, (11) Dates of observation for radial velocity data in Heliocentric Julian Date minus 2,458,000. }
\end{deluxetable*}

\clearpage

\movetabledown=0.7in
\begin{longrotatetable}
\begin{deluxetable*}{rcrrrccrrrrrrrcr}
\tablecaption{Astrometric Data for Selected SBN Central Stars \label{tab:BSNkinematics}}
\tabletypesize{\scriptsize}
\tablewidth{0pt}
\tablehead{
\colhead{SBN ID}&\colhead{Alias}&\colhead{$\alpha$ (2000)}&\colhead{$\delta$ (2000)}&\colhead{EDR3 ID}&\colhead{G}     &\colhead{Env.}&\colhead{$\pi$}  &\colhead{$\sigma_{\pi}$}&\colhead{$\mu_\alpha$}	&\colhead{$\sigma_{\mu_\alpha}$}&\colhead{$\mu_\delta$}   &\colhead{$\sigma_{\mu_\delta}$}&\colhead{Code}&\colhead{Sp.T.}&\colhead{Ref.} \\
\colhead{}      &\colhead{}     &\colhead{(deg.)}         &\colhead{(deg.)}         &\colhead{}       &\colhead{(mag.)}&\colhead{}    &\colhead{(mas)}&\colhead{(mas)}	      &\colhead{(mas yr$^{-1}$)}&\colhead{(mas yr$^{-1}$)}	&\colhead{(mas yr$^{-1}$)}&\colhead{(mas yr$^{-1}$)}	  &\colhead{}	 &\colhead{}	   &\colhead{}     \\	
\colhead{(1)}   &\colhead{(2)}  &\colhead{(3)}            &\colhead{(4)}	    &\colhead{(5)}    &\colhead{(6)}   &\colhead{(7)} &\colhead{(8)}  &\colhead{(9)}	      &\colhead{(10)}		&\colhead{(11)} 		&\colhead{(12)} 	  &\colhead{(13)}		  &\colhead{(14)}&\colhead{(15)}   &\colhead{(16)} 
}
\startdata
  1 &          \nodata & 267.0320 & -29.1321 & 4057291747277127040 & 10.73 &     I & 0.3510 & 0.0180 &   1.2820 & 0.0180 &  -1.2100 & 0.0110 & 3 & 	 (OB+)       & S71     \\
  2 &          \nodata & 267.6164 & -29.2128 & 4057244154701712000 & 15.75 &     I & 0.4600 & 0.0650 &  -1.0050 & 0.0670 &  -2.3060 & 0.0410 & 3 & 	 \nodata     & \nodata \\
  3 &          \nodata & 267.1745 & -28.1110 & 4057634103373474560 & 16.29 &     H & 0.6370 & 0.0520 &  -2.2260 & 0.0610 & -12.1090 & 0.0370 & 3 & 	 B           & \nodata \\
  7 &          \nodata & 269.6277 & -26.1637 & 4064254301560852352 & 12.74 &    FB & 0.8220 & 0.0170 &   1.9260 & 0.0200 &  -1.8690 & 0.0140 & 3 & 	 \nodata     & \nodata \\
 10 &        HD 314937 & 270.0728 & -25.2374 & 4064417063652618752 & 10.62 &     I & 0.4060 & 0.0170 &   0.8920 & 0.0160 &  -2.1000 & 0.0120 & 3 & 	 (B)         & N95     \\
 12 &          \nodata & 270.2754 & -24.3596 & 4069013297115920256 & 13.52 &     I & 0.2680 & 0.0200 &  -0.1330 & 0.0210 &  -2.5960 & 0.0150 & 3 & 	 OB          & \nodata \\
 14 &          \nodata & 270.1691 & -23.6141 & 4069123317001611648 & 16.75 &     H & 0.3300 & 0.0790 &  -0.6790 & 0.0730 &  -1.9400 & 0.0510 & 3 & 	 \nodata     & \nodata \\
 16 &          \nodata & 270.2301 & -22.9603 & 4069437463777653632 & 11.68 &     I & 0.3260 & 0.0200 &   0.1280 & 0.0190 &  -0.9850 & 0.0140 & 3 & 	 B           & \nodata \\
\enddata
\tablecomments{References for source of spectral types in Column 18: A16--\citet{Alexander2016}; B96--\citet{Barnbaum1996}; B18--\citet{Berlanas2018};  
C93--\citet{Cannon1993}; G11--\citet{Gvaramadze2011}; H56--\citet{Hiltner1956}; H64--\citet{Hardorp1964}; H93--\citet{Hillenbrand1993};
H75--\citet{Houk1975}; H78--\citet{Houk1978}; H88--Houk1988); K10--\citet{Kobulnicky2010}; M16--\citet{Maiz2016}; M54--\citet{Munch1954}; 
 M95--\citet{Massey1995}; N95--\citet{Nesterov1995};
P12--\citet{Parthasarathy2012}; R03--\citet{Reed2003}; R19--\citet{Roman-Lopes2019}; S11--\citet{Sota2011}; S14--\citet{Sota2014}; S17--\citet{Sana2017}; 
S60--\citet{Stock1960}; S71--\citet{Stephenson1971}; V93--\citet{Vijapurkar1993}; W07--\citet{Wolff2007}; W71--\citet{Walborn1971}
}
\end{deluxetable*}
\end{longrotatetable}

\clearpage

\startlongtable
\begin{deluxetable*}{rrrrrrrrrrr}
\tablecaption{Kinematic Data for SBN Central Stars \label{tab:BSNkinematics2}}
\tabletypesize{\scriptsize}
\tablewidth{0pt}
\tablehead{
\colhead{SBN ID}&\colhead{\vtwod}&\colhead{$\sigma_{\rm 2D}$}&\colhead{\vthreed}&\colhead{$\sigma_{\rm 3D}$}&\colhead{$PA_{\rm m}$}&\colhead{$PA_{\rm k}$}&\colhead{$\sigma_{PA_k}$}&\colhead{$\Delta_{PA(m-k)}$}&\colhead{$\sigma_{\Delta}$} & \colhead{Code} \\
\colhead{}      &\colhead{(\kms)}&\colhead{(\kms)}           &\colhead{(\kms)}  &\colhead{(\kms)}           &\colhead{(deg.)}      &\colhead{(deg.)}      &\colhead{(deg.)}         &\colhead{(deg.)}           &\colhead{(deg.)}            &\colhead{}  \\   
\colhead{(1)}   &\colhead{(2)}   &\colhead{(3)}              &\colhead{(4)}     &\colhead{(5)}              &\colhead{(6)}         &\colhead{(7)}         &\colhead{(8)}            &\colhead{(9)}	        &\colhead{(10)}	             &\colhead{(11)} \\ 
}
\startdata
  1 &  19 &   2 & -999 & -999 &  27 &  89 &   9 &  -62 &  12 & 3 \\ 
  2 &  19 &   4 & -999 & -999 & 145 & 236 &  11 &  -91 &  14 & 3 \\ 
  3 &  83 &   8 & -999 & -999 & 200 & 193 &   1 &    7 &   8 & 3 \\ 
  7 &  11 &   1 & -999 & -999 &  25 &  63 &  11 &  -38 &  14 & 3 \\ 
 10 &  19 &   2 & -999 & -999 & 140 & 130 &  13 &   10 &  15 & 3 \\ 
 12 &  32 &   3 & -999 & -999 & 180 & 182 &   4 &   -2 &   9 & 3 \\ 
 14 &  20 &  33 & -999 & -999 & 210 & 225 &  22 &  -15 &  24 & 3 \\ 
 16 &   5 &   2 & -999 & -999 &  65 &  27 &  41 &   38 &  42 & 3 \\
\enddata
\end{deluxetable*}

\clearpage

\begin{deluxetable*}{lr}
\tablecaption{Summary of Principal Conclusions \label{tab:summary}}
\tablehead{
\colhead{Property}&\colhead{Value} }
\startdata
\cutinhead{Galactic O Star Catalog Findings}
Overall Runaway Fraction & 22$^{+3}_{-3}$\ \\
Multiplicity fraction among 25 runaways  & $>$28\% \\
16th/50th/84th percentiles of \vtwod\   &  5.6/11.4/32.6 \kms \\
\cutinhead{Bowshock Nebula Central Star Findings}
Runaway Fraction & 24$^{+9}_{-7}$\% \\
Multiplicity Fraction & $>$36\%$\pm$6\% \\
16th/50th/84th percentiles of \vtwod\    &  6.5/14.6/32.6 \kms  \\
\enddata
\end{deluxetable*}

\bibliography{ms}

\end{document}